\documentclass[11pt]{article}
\usepackage[backend=biber,style=apa, natbib=true, sortcites=false, uniquename=false]{biblatex}
\usepackage{color}
\usepackage[dvipsnames]{xcolor}
\usepackage{epsfig,lscape}
\usepackage{graphicx} % Required for inserting images
\usepackage{tikz}
\usepackage{amssymb,amsfonts,amsmath,amsthm}
\usepackage{rotating}
\usepackage{bm}
\usepackage{bbm}
\usepackage{mathtools}
\usepackage{threeparttable}
\usepackage{booktabs}
\usepackage{caption}
\usepackage{subcaption}
\usepackage{algorithm}
\usepackage[noend]{algorithmic}

\usepackage{enumitem}

\definecolor{darkblue}{RGB}{15,20,150}
\usepackage[colorlinks=true,linkcolor=darkblue,citecolor=darkblue,urlcolor=blue,]{hyperref}

\usepackage{titlesec}
\titleformat*{\section}{\large \bfseries\color{darkblue}}
\titleformat*{\subsection}{\large\bfseries\color{darkblue}}
\titleformat*{\subsubsection}{\normalsize\bfseries\color{darkblue}}
\titleformat*{\paragraph}{\normalsize\bfseries\color{darkblue}}

\usepackage[left=1.1in,right=1.1in,top=.9in,bottom=0.9in]{geometry}

\allowdisplaybreaks
\newtheorem{pro}{Proposition}

\newtheorem{ass}{Assumption}
\newtheorem{rem}{Remark}

\theoremstyle{definition}

\def\me{\mathrm e}

\def\dif{\mathrm d}
\def\goto{\rightarrow}

\def\diag{\mathrm{diag}}

\def\bbR{\mathbb R}

\def\bbeta{\boldsymbol{\beta}}
\def\btau{\boldsymbol{\tau}}
\def\btheta{\boldsymbol{\theta}}
\def\bmu{\boldsymbol{\mu}}
\def\bepsilon{\boldsymbol{\epsilon}}
\def\bXi{\boldsymbol{\Xi}}
\def\bTheta{\boldsymbol{\Theta}}

\def\ba{\mathbf{a}}

\def\bg{\mathbf{g}}
\def\bh{\mathbf{h}}
\def\bs{\mathbf{s}}
\def\bt{\mathbf{t}}
\def\bu{\mathbf{u}}
\def\bv{\mathbf{v}}
\def\bw{\mathbf{w}}
\def\bx{\mathbf{x}}
\def\by{\mathbf{y}}
\def\bz{\mathbf{z}}
\def\bzero{\mathbf{0}}
\def\bA{\mathbf{A}}
\def\bB{\mathbf{B}}
\def\bC{\mathbf{C}}
\def\bD{\mathbf{D}}
\def\bF{\mathbf{F}}
\def\bG{\mathbf{G}}
\def\bI{\mathbf{I}}
\def\bQ{\mathbf{Q}}
\def\bR{\mathbf{R}}
\def\bT{\mathbf{T}}

\def\bV{\mathbf{V}}
\def\bW{\mathbf{W}}
\def\bX{\mathbf{X}}

\def\bZ{\mathbf{Z}}

\def\bbE{\mathbb{E}}

\def\cD{\mathcal{D}}

\def\cS{\mathcal{S}}
\def\cT{\mathcal{T}}
\def\cU{\mathcal{U}}
\def\cV{\mathcal{V}}

\def\cZ{\mathcal{Z}}

\title{\textcolor{darkblue}{Nearest-Neighbor Non-Gaussian Processes for Geostatistical Modeling}}
\author{Penghui Fu, Yuhan Dong, Jianhua Z. Huang\thanks{Corresponding author: \href{mailto:jhuang@cuhk.edu.cn}{jhuang@cuhk.edu.cn}}}
\date{%
    \vspace{-.1in}
    School of Data Science \\
    The Chinese University of Hong Kong, Shenzhen\\[2ex]
    \today 
}

\begin{document}
\maketitle
%\tableofcontents

\begin{abstract}
We develop a general class of nearest-neighbor non-Gaussian processes (NNnGP) for modeling geostatistical data. 
By introducing non-linear conditional mean functions into the Vecchia approximation, NNnGP extends the popular nearest-neighbor Gaussian process (NNGP) to effectively capture complex, non-Gaussian spatial characteristics.
To ensure coherent spatial predictions from a single realization, we propose a regularized homogeneity condition across the univariate conditionals. We then formulate a Bayesian non-parametric construction of the conditional mean using Gaussian processes and embed the resulting NNnGP as a non-Gaussian spatial prior within a hierarchical regression model. 
For posterior computation and prediction, we adopt a normalizing flow-based variational inference approach. 
Numerical experiments on synthetic data and real-world PRISM precipitation anomalies demonstrate that NNnGP captures local non-linear dependencies and significantly mitigates the underestimation of extreme spatial events, all the while maintaining global predictive accuracy.
\end{abstract}


%\tableofcontents

\section{Introduction}

Gaussian processes (GPs) have long served as the workhorse for spatial modeling due to their mathematical tractability and natural uncertainty quantification \parencite{stein1999interpolation,cressie2015statistics,banerjee2025hierarchical}. 
However, the ubiquitous application of GPs is fundamentally constrained by the underlying Gaussianity assumption, which renders traditional GPs inadequate for capturing real-world data exhibiting non-Gaussian characteristics, such as skewness, heavy tails, and extremes with complex dependence structures 
\parencite{padoan2010likelihood,davison2012statistical,sun2015stochastic,xu2017tukey}.

Several approaches have been proposed for modeling non-Gaussian random fields. 
A common practice is to transform the target process before fitting a GP \parencite{de1997bayesian,rios2019compositionally,lin2020transformation}, or inversely, to model the process of interest as a transformation of an underlying GP \parencite{lazaro2012bayesian,xu2017tukey,yan2020multivariate}.
However, monotonic transformations of GPs preserve the restrictive Gaussian copula, failing to capture tail dependence. 
Spatial copula models address this by directly coupling pre-specified marginals with non-Gaussian copulas \parencite{krupskii2018factor,krupskii2019copula,beck2020predicting}.

Alternatively, non-Gaussian distributions can be modeled as a mixture of Gaussian distributions, such as location-scale mixtures \parencite{zareifard2018modeling,tagle2020hierarchical,bevilacqua2021non} or Dirichlet process mixtures \parencite{gelfand2005bayesian,duan2007generalized,wade2025bayesian}.
The mixture components themselves can also be non-Gaussian \parencite{zheng2023nearest}.
Another line of work constructs non-Gaussian spatial Mat\'ern fields as solutions to stochastic partial differential equations (SPDEs) with non-Gaussian noise \parencite{bolin2014spatial, lindgren2022spde}. 

For non-Gaussian responses (e.g., binary, count, or nonnegative data), generalized linear models (GLMs) are widely used---often referred to as spatial generalized linear mixed models in statistics \parencite{diggle1998model} or generalized GPs in machine learning \parencite{chan2011generalized}. 
These models assume that, conditional on a latent random field, the observations are independent with marginals from the exponential family. 
The conditional mean of the response is connected to the latent field via a link function.
Notably, the underlying latent field in GLM-type models usually remains Gaussian, which conceptually distinguishes them from methods that directly model non-Gaussian continuous processes.

In this article, we propose a flexible class of non-Gaussian random processes.
Our construction follows a two-step procedure. 
First, we specify a multivariate distribution over a set of reference locations.
To reduce modeling complexity, we introduce conditional independence by factorizing this joint distribution into a product of univariate conditionals over a directed acyclic graph.
Second, we extend this distribution to the full continuous domain by assuming location-wise conditional independence given the reference set.
The two steps combined define a valid random process.
Importantly, we model each conditional as a Gaussian distribution whose mean is a 
nonlinear function of its neighbors.
This non-linearity empowers the model to capture complex local dependencies, yielding a non-Gaussian process. 
Furthermore, to ensure consistent parameter estimation and coherent kriging from a single realization, we introduce a regularized homogeneity condition on the conditional means and variances across locations. 
The resulting model encompasses the popular nearest-neighbor Gaussian process \parencite[NNGP,][]{datta2016hierarchical} as a special case, and is referred to as the nearest-neighbor non-Gaussian process (NNnGP).

To realize the NNnGP framework, we develop a Bayesian non-parametric formulation for the conditional mean function that takes an additive form comprising a linear and a non-linear term. 
Inspired by recent advances in Bayesian generative modeling for spatial data \parencite{katzfuss2024scalable}, we model the non-linear component using a GP, explicitly enforcing the screening effect and other desirable properties to improve model interpretability. 
We then embed this NNnGP as a non-Gaussian prior for the latent spatial field within a hierarchical regression model. 
To address computational bottlenecks, we leverage low-rank approximations for the GP component and adopt a normalizing flow-based variational inference (VI) approach for posterior computation and spatial prediction.

The remainder of the article is organized as follows.
Section~\ref{sec:NNnGP-general} formally introduces the NNnGP modeling framework. 
Section~\ref{sec:NNnGP-GP} details the Bayesian non-parametric formulation of the conditional mean.
Section~\ref{sec:illustrations} conducts numerical studies on both synthetic and real-world datasets.
Section~\ref{sec:conclusion} concludes the paper with a summary and discussion of future work.
Proofs and additional technical details are provided in the Supplementary Materials.

\section{Nearest-Neighbor Non-Gaussian Processes}\label{sec:NNnGP-general}
\subsection{Model Formulation}
We begin by formalizing the general class of NNnGP models. 
Let $w(\bs)$ denote a random process defined over a domain $\cD\subset \bbR^d$.
We introduce a set of $k$ distinct reference locations, denoted and ordered as $\cS=\{\bs_1,\ldots,\bs_k\}$.\footnote{The reference points need not coincide with or be a part of the observed locations.}
Using the chain rule of probability, the joint density of the process at these reference locations, $\bw_\cS = (w(\bs_1),\ldots,w(\bs_k))^\top$, can be decomposed as 
\begin{equation}\label{eq:autoregressive-model}
    p(\bw_{\cS}) = \prod_{i=1}^k p(w(\bs_i)|w(\bs_{i-1}),\ldots,w(\bs_1)).
\end{equation}
As $i$ increases, the conditioning set on the right-hand side of (\ref{eq:autoregressive-model}) grows prohibitively large.
To retain computational tractability, we employ the Vecchia approximation 
\parencite{vecchia1988estimation, katzfuss2021general}, which replaces
the conditioning sets on the right-hand side of (\ref{eq:autoregressive-model}) with smaller ones of size at most $m\ll k$.
Specifically, we define the neighbor set $N(\bs_1) = \varnothing$, and $N(\bs_i)= \cS_i=\{\bs_1,\ldots,\bs_{i-1}\}$ for $i=2,\ldots,m$. 
For subsequent locations $\bs_i\in\{\bs_{m+1},\ldots,\bs_k\}$, $N(\bs_i)\subseteq \cS_i$ is chosen as the $m$-nearest neighbors of $\bs_i$ among the previously ordered locations $\cS_i$
with respect to the Euclidean distance.
We assume that for $i=1,\ldots,k$,
\begin{equation}\label{eq:vechia-reference}
    p(w(\bs_i)|w(\bs_{i-1}),\ldots,w(\bs_1)) = p(w(\bs_i)|\bw_{N(\bs_i)}).
\end{equation}
That is, $w(\bs_i)$ is conditionally independent of $\bw_{\cS_i\backslash N(\bs_i)}$ given $\bw_{N(\bs_i)}$. 
The conditional independence structure of $\bw_\cS$ can be represented by a directed acyclic graph \parencite[DAG,][]{jordan2004graphical}, where $\cS$ is viewed as the set of nodes, and for every pair of nodes $(\bs_j, \bs_i)$, there is a directed edge from $\bs_j$ to $\bs_i$ if and only if $\bs_j\in N(\bs_i)$.

At the core of our non-Gaussian extension is the assumption that each univariate conditional distribution remains Gaussian, but with a potentially non-linear mean:
\begin{equation}\label{eq:conditional-reference}
    p(w(\bs_i)|\bw_{N(\bs_i)}) = N(w(\bs_i)|f_i(\bw_{N(\bs_i)}, N(\bs_i),\bs_i), \sigma_i^2(N(\bs_i),\bs_i)).
\end{equation}
Here, $f_i \colon \bbR^{m\wedge (i-1)}\times \cD^{m\wedge (i-1)}\times \cD \goto \bbR$ and $\sigma_i^2 \colon \cD^{m\wedge (i-1)}\times\cD \goto \bbR_{\geq 0}$ are the conditional mean and variance functions, respectively; 
$N(\bs_i)$ as an argument of $f_i$ and $\sigma^2_i$ is understood as an ordered vector of locations.
Similarly, $\bw_{N(\bs_i)}$ is a vector of realizations following the same order as $N(\bs_i)$.
We set $f_1\equiv 0$, and $\sigma_1^2$ as a constant.
By definition, the $i$-th conditional mean $f_i(\bw_{N(\bs_i)}, N(\bs_i),\bs_i)$ depends on both the neighbor locations and their actual realizations, while the conditional variance $\sigma_i^2(N(\bs_i),\bs_i))$ depends only on the spatial locations.

To generalize this distribution from the reference set $\cS$ to the entire domain $\cD$, we consider an arbitrary finite set $\cU=\{\bu_1,\ldots,\bu_r\}$ such that $\cS\cap\cU = \varnothing$.
Let $\bw_{\cU} = (w(\bu_1),\ldots,w(\bu_r))^\top$ and denote $N(\bu_i)\subseteq \cS$ as the $m$-nearest neighbor set for $\bu_i$ within $\cS$. 
Conditional on the realizations at the reference locations, we assume that $\bw_{\cU}$ are location-wise independent, adhering to a Vecchia approximation similar to (\ref{eq:vechia-reference}): 
\begin{equation}\label{eq:vecchia-non-reference}
    p(\bw_{\cU}|\bw_{\cS}) = \prod_{i=1}^r p(w(\bu_i)|\bw_{\cS}) = \prod_{i=1}^r p(w(\bu_i)|\bw_{N(\bu_i)}),
\end{equation}
where each conditional is again Gaussian, with a common conditional mean function $f_*$ and variance function $\sigma^2_*$:
\begin{equation}\label{eq:conditional-non-reference}
     p(w(\bu_i)|\bw_{N(\bu_i)}) = N(w(\bu_i)|f_*(\bw_{N(\bu_i)}, N(\bu_i),\bu_i), \sigma_*^2(N(\bu_i),\bu_i)).
\end{equation}
% A similar independence assumption, namely the Fully Independent Conditional (FIC) assumption, was used for Gaussian process approximate inference. In that context, $\cS$ are called \textit{inducing points} \parencite{quinonero2005unifying}. 
% To make (\ref{eq:conditional-non-reference}) valid, the reference points $\cS$ need to be dense in $\cD$, while for GP approximate inference, the number of inducing points is typically small.

Combining (\ref{eq:autoregressive-model})--(\ref{eq:conditional-non-reference}) defines a collection of probability distributions for all finite subsets in $\cD$. 
Let $\cV\subset\cD$ be a finite set. We have
\begin{equation*}
    p(\bw_{\cV}) = \int p(\bw_{\cV\cup\cS})\,\dif \bw_{\cS\setminus\cV} =  \int p(\bw_{\cS})p(\bw_{\cV\setminus\cS}|\bw_{\cS})\,\dif \bw_{\cS\setminus\cV},
\end{equation*}
where $p(\bw_{\cS})$ is determined by (\ref{eq:autoregressive-model})--(\ref{eq:conditional-reference}), and $p(\bw_{\cV\setminus\cS}|\bw_{\cS})$ is given by (\ref{eq:vecchia-non-reference})--(\ref{eq:conditional-non-reference}).
These distributions naturally satisfy the Kolmogorov's consistency condition; thereby corresponding to a valid spatial process over $\cD$, which is generally non-Gaussian. 
In fact, by the property of multivariate Gaussian distributions, $\bw_\cS$ is jointly Gaussian if and only if the $i$-th conditional mean $f_i$ is linear with respect to $\bw_{N(\bs_i)}$ for all $i=1,\ldots,k$.

\subsection{The Regularized Homogeneity Assumption}
% From a generative perspective, the spatial process $w$ can be simulated as follows. 
% Let $\epsilon(\cdot)$ be a Gaussian white-noise process on $\cD$ with unit variance. 
% Then, $w$ can be simulated by the following two-step procedure.
% \begin{enumerate}
%     \item Recursively simulate $w(\bs_1),\ldots,w(\bs_k)$ by
%     $$
%     w(\bs_i) = f_i(\bw_{N(\bs_i)}, N(\bs_i),\bs_i) + \sigma_i(N(\bs_i),\bs_i)\cdot\epsilon(\bs_i), \quad \text{for }i=1,\ldots,k.
%     $$
%     \item For $\cU\subset\cD\backslash\cS$, simulate for each $\bu\in\cU$ independently by
%     $$
%     w(\bu) = f_*(\bw_{N(\bu)}, N(\bu),\bu) + \sigma_*(N(\bu),\bu)\cdot\epsilon(\bu).
%     $$
% \end{enumerate}

% The above procedure can be viewed as a sequence of spatial extrapolations, where we recursively extrapolate the realization at each reference point from the realizations at previous reference points, then use these to extrapolate outside $\cS$.
% % The procedure of simulating $\bw_\cS$ can also be viewed as an autoregressive flow with a single affine autoregressive layer. 
% % See Section 3.1.3. in \textcite{papamakarios2021normalizing} for related discussions.
% The conditional mean functions $f_i$ and $f_*$ encode how the extrapolation is performed within and outside $\cS$, respectively. 
% Similarly, the variance functions $\sigma_i^2$ and $\sigma_*^2$ encode the uncertainty of extrapolation within and outside $\cS$, respectively. 

Geostatistical modeling typically requires performing spatial interpolation (kriging) from a single, often noisy, realization of the spatial field $w$.
Doing so inherently requires an information-sharing mechanism across the domain. 
To this end, we impose a regularized homogeneity assumption:
\begin{ass}\label{ass:homogeneity}
    The mean and variance functions are continuous and homogeneous. That is, there exist two continuous functions $f \colon \bbR^m\times \cD^{m}\times \cD \goto \bbR$ and $\sigma^2 \colon \cD^{m}\times\cD \goto \bbR_{\geq 0}$, such that
    \begin{align*}
        &f_{m+1} = f_{m+2} = \cdots = f_k = f_* = f, \\
        &\sigma_{m+1}^2 = \sigma_{m+2}^2 = \cdots = \sigma_k^2 = \sigma_*^2 = \sigma^2.
    \end{align*}
    As a result, (\ref{eq:conditional-reference}) and (\ref{eq:conditional-non-reference}) can be unified into
\begin{equation}\label{eq:conditional-unified}
    p(w(\bv)|\bw_{N(\bv)}) = N(w(\bv)|f(\bw_{N(\bv)}, N(\bv),\bv), \sigma^2(N(\bv),\bv)),
\end{equation}
for a generic query location $\bv\in\cD\backslash\cS_{m+1}$.
\end{ass}

Assumption~\ref{ass:homogeneity} is fundamental for our framework.
It enforces a universal conditional rule across all locations—both within and outside the reference set.
However, homogeneity alone, in terms of a common conditional mean and variance function, is too general and fails to provide a useful information-sharing mechanism. 
Without regularization, a highly flexible function $f$ over $\bbR^m\times \cD^{m}\times \cD$ could interpolate essentially arbitrary conditional behaviors at finite spatial coordinates. 
By enforcing continuity, we constrain the spatial transition of the conditional distributions, providing the necessary regularization.
In Section~\ref{sec:NNnGP-GP}, we study a specific parameterization of $f$ and $\sigma^2$, which provides further regularization.

Conceptually, Assumption~\ref{ass:homogeneity} plays a role analogous to stationarity by sharing statistical dependence across space. 
However, it is fundamentally distinct from classical strict or weak stationarity. Because the finite-dimensional joint distribution of $w$ intrinsically depends on the explicit locations and the specific ordering of the reference points, the resulting process is not translation invariant and thus non-stationary in the traditional sense.

\subsection{Related Models}\label{sec:related models}
\subsubsection{NNGP}
The Nearest-Neighbor Gaussian Process \parencite[NNGP,][]{datta2016hierarchical} is a scalable GP model for large-scale geostatistical modeling. 
Let $p_\text{parent}$ denote the distribution of a parent GP with mean $0$ and covariance function $C$. 
For the reference locations, NNGP assumes that
\begin{equation}\label{eq:vecchia-NNGP}
    p(w(\bs_i)|w(\bs_{i-1}),\ldots,w(\bs_1)) = p_\text{parent}(w(\bs_i)|\bw_{N(\bs_i)}) = N(w(\bs_i)|\bB_{\bs_i,N(\bs_i)}\bw_{N(\bs_i)}, C_{\bs_i|N(\bs_i)}),
\end{equation}
where $\bB_{\bs_i,N(\bs_i)} = \bC_{\bs_i,N(\bs_i)}\bC_{N(\bs_i)}^{-1}$, $C_{\bs_i|N(\bs_i)}=C(\bs_i,\bs_i)-\bC_{\bs_i,N(\bs_i)}\bC_{N(\bs_i)}^{-1}\bC_{N(\bs_i),\bs_i}$, $\bC_{N(\bs_i)}$ is the covariance matrix of $\bw_{N(\bs_i)}$ under $p_\text{parent}$, and $\bC_{N(\bs_i),\bs_i}$ is the cross-covariance matrix of $\bw_{N(\bs_i)}$ and $w(\bs_i)$ under $p_\text{parent}$.
For non-reference locations $\cU=\{\bu_1,\ldots,\bu_r\}\subset \cD\backslash\cS$, NNGP assumes 
\begin{equation}\label{eq:wu-given-ws-NNGP}
    p(\bw_{\cU}|\bw_{\cS}) = \prod_{i=1}^r p_\text{parent}(w(\bu_i)|\bw_{N(\bu_i)}) = \prod_{i=1}^r N(w(\bu_i)|\bB_{\bu_i,N(\bu_i)}\bw_{N(\bu_i)}, \bC_{\bu_i|N(\bu_i)}),
\end{equation}
where $\bB_{\bu_i,N(\bu_i)}$ and $\bC_{\bu_i|N(\bu_i)}$ are defined analogously to
(\ref{eq:vecchia-NNGP}).

An inspection of (\ref{eq:vecchia-NNGP}) and (\ref{eq:wu-given-ws-NNGP}) reveals that NNGP is a special case of NNnGP and the Assumption~\ref{ass:homogeneity} holds, provided that the neighbor set size is fixed at $m$ (except for the first $m$ reference points), with the mean and variance function given by
\begin{equation}\label{eq:mean-variance-NNGP}
    f(\bw, N, \bv) = \bB_{\bv,N}\bw,\quad \sigma^2(N,\bv) = \bC_{\bv|N},
\end{equation}
respectively. 
Here, $\{\bw, N, \bv\}$ denotes a generic tuple of $m$ (neighbor) realizations, $m$ (neighbor) locations, and one (query) location.

\subsubsection{NNMP}
The Nearest-Neighbor Mixture Model \parencite[NNMP,][]{zheng2023nearest} also adopts a similar two-step procedure.
The key difference lies in how the conditionals are modeled. 
For a generic query location $\bv\in\cD\backslash\cS_{m+1}$, let $N(\bv)=\{\bv_{(1)},\ldots,\bv_{(m)}\}$ denote the $m$ neighbor locations in ascending order with respect to their distance to $\bv$.
NNMP introduces non-Gaussianity through a spatially varying mixed conditional distribution
\begin{equation}\label{eq:conditional-nnmp}
    p(w(\bv)|\bw_{N(\bv)}) = \sum_{j=1}^m \omega_j(\bv) p_{\bv,j}(w(\bv)|w(\bv_{(j)})),
\end{equation}
where $\omega_1(\bv),\ldots,\omega_m(\bv)$ are $m$ non-negative spatially varying weights summing up to $1$, and $\{p_{\bv,j}(\cdot|\cdot)\}_{j=1}^m$ are a collection of spatially varying conditional distributions.
The $j$-th weight $\omega_j(\bv)$ and conditional distribution $p_{\bv,j}$ in general depend on the exact location of the $j$-th neighbor $\bv_{(j)}$, which is suppressed for brevity.

The special bivariate structure in the mixture components facilitates the modeling and analysis of NNMP.
Under some conditions on $\{p_{\bv,j}\}_{j=1}^m$, NNMPs have stationary marginals, which can accommodate generic, even discrete, distributions.
In addition, the (bivariate) tail dependence of $\{p_{\bv,j}\}_{j=1}^m$ quantifies the multivariate tail dependence of $(w(\bv)$, $w(\bv_{(1)})$, $\ldots$, $w(\bv_{(m)}))$.
These two properties, in turn, guide the formulation of $\{p_{\bv,j}\}_{j=1}^m$ and lead to various instances of NNMPs. 
Despite these appealing theoretical properties, NNMPs are generally less scalable than NNGP when embedded as latent spatial fields in hierarchical models. 
Posterior computation in NNMP can be prohibitive, except for the regression model with the Gaussian NNMP, whose mixture components are conditionals of bivariate Gaussians. 
See Appendix E.1 of the supplementary material of \textcite{zheng2023nearest} for related discussion.

% The simplified structure of the mixture in (\ref{eq:conditional-nnmp}) facilitates the modeling of NNMP. 
% The group of spatially varying densities $\{p_{\bv,j}\}_{j=1}^m$ are determined by a group of bivariate random vectors $\{(U_{\bv,j},V_{\bv,j})\}$, such that $p_{\bv,j}$ is the conditional density of $U_{\bv,j}$ given $V_{\bv,j}$.
% To reduce the modeling complexity, NNMP builds $\{(U_{\bv,j}, V_{\bv,j})\}$ as "copies" of base bivariate vectors $\{(U_j, V_j)\}$, whose parameters are extended to be spatially varying, i.e., depending on $\bv$ and $\bv_{(j)}$ for each $j$.
% The gain is that $\{(U_{\bv,j},V_{\bv,j})\}$ can be built by separately modeling the base bivariate vectors $\{(U_j, V_j)\}$ and modeling how their parameters vary in space.
% The base bivariate distributions can be modeled as a location-scale mixture of bivariate Gaussian distributions, or via bivariate copulas.

% The simplified structure of the mixture in (\ref{eq:conditional-nnmp}) also facilitates the analysis of NNMP. 
% Remarkably, under suitable conditions, NNMP has stationary marginals, which can be flexibly modeled as any generic (including discrete) distribution, and the multivariate tail dependence of $(w(\bv)$, $w(\bv_{(1)})$, $\ldots$, $w(\bv_{(1)}))$ is shown to be related to the tail dependence of the bivariate distributions of $\{(U_j, V_j)\}$.
% See \textcite{zheng2023nearest} for more details.

\section{A Bayesian Non-Parametric Formulation for The Conditional Mean}\label{sec:NNnGP-GP}
The NNnGP model hinges on the conditional mean $f$ and variance $\sigma^2$.
In this section, we propose a Bayesian non-parametric formulation of $f$ and a parametric formulation of $\sigma^2$, designed to strike a balance among model expressivity, estimation efficiency, and interpretability.
% \footnote{As discussed in Section~\ref{sec:homogeneity-general-NNnGP}, deep learning methods, such as GNNs, may also be considered.}
Our formulation of $f$ is inspired by the Bayesian transport map recently proposed by \textcite{katzfuss2024scalable}.
Nonetheless, a critical difference is that our NNnGP formally defines a valid continuous stochastic process over the entire spatial domain, while their approach learns a multivariate distribution over a \textit{fixed} set of locations (primarily for generative modeling).
Modeling a full spatial process enables coherent spatial predictions at unobserved locations based on a single realization, which purely generative models cannot fulfill.

\subsection{Formulations of the Conditional Mean and Variance}\label{sec:formulation-mean-variance}
Before presenting the exact formulation, we first discuss several desired properties of $f$ and $\sigma^2$ that will improve model interpretability.

\begin{itemize}[noitemsep]
\item \textbf{Non-linearity} \quad For NNGP, the conditional mean function $f(\bw,N,\bv)$ in (\ref{eq:mean-variance-NNGP}) is linear in $\bw$, which implies that $(\bw_\cS,\bw_\cU)$ is jointly Gaussian for any finite $\cU\subset\cD\backslash\cS$. 
Therefore, NNGP is still a Gaussian process.
To introduce the non-Gaussianity, it suffices to allow $f(\bw, N,\bv)$ to be nonlinear in terms of $\bw$.

\item \textbf{Screening effect}\quad In spatial statistics, it is often noted that when predicting at a location $\bv$, observations close to $\bv$ tend to have a greater impact than observations that are far away. 
This phenomenon is called the screening effect \parencite{stein2002screening}. 
To be consistent with the screening effect, the dependency of $f(\bw,N,\bv)$ on a neighbor realization in $\bw$ should decay with its distance to $\bv$. 

\item \textbf{Continuity in extrapolation}\quad 
Given $\bw_N$, the conditional mean and variance of $w(\bv)$ is given by 
$f(\bw_N,N,\bv)$ and $\sigma^2(N,\bv)$, respectively, which should become
degenerate if $\bv$ is very close to $N$.
To be specific, let $\bu\in N$ be a neighbor location in $N$. 
As $\bv\goto\bu $, $f(\bw_N,N,\bv)$ should converge to $w(\bu)$, and the uncertainty $\sigma^2(N,\bv)$ should vanish to $0$.

\item \textbf{Translation invariance} \quad To reduce model complexity, following a similar notion to stationarity, we may assume that $f$ and $\sigma$ are both translation invariant in terms of locations. That is, for any spatial shift $\Delta\bv\in\bbR^d$, it holds that
\begin{equation*}
    f(\bw, N + \Delta\bv, \bv + \Delta\bv) = f(\bw, N, \bv),\quad \sigma(N + \Delta\bv, \bv + \Delta\bv) = \sigma(N, \bv),
\end{equation*}
where $N + \Delta\bv$ is defined location-wise. 
In other words, the conditional mean and variance only depend on the \textit{relative} locations of $\bv$ compared with its neighbors. 

\item \textbf{Permutation invariance}\quad In (\ref{eq:conditional-unified}), the neighbor locations $N(\bv)$ and their realizations $\bw_{N(\bv)}$ are ordered before being fed into the mean and variance functions. 
For better interpretation, we would like $f$ and $\sigma$ to satisfy permutation invariance with respect to locations. That is,
\begin{equation*}
    f(\bw^\dag, N^\dag, \bv) = f(\bw, N, \bv),\quad \sigma(N^\dag, \bv) = \sigma(N, \bv),
\end{equation*}
where $N^\dag$ is $N$ with locations permutated, and $\bw^\dag$ is $\bw$ with the same permutation. 
Let $\bw=(w_1,\ldots,w_m)^\top$ and $N=(\bu_1,\ldots,\bu_m)^\top$.
It amounts to require that $f(\bw,N,\bv)$ (resp., $\sigma(N,\bv)$) is a set function of $\{w_i, \bu_i\}_{i=1}^m$ (resp., $\{\bu_i\}_{i=1}^m$), for a fixed $\bv$.
\end{itemize}

Motivated by the desired properties above, we propose the following formulations of $f$ and $\sigma$.
Let $C$ be a Mat\'ern covariance function with smoothness $3/2$, global variance $\sigma^2_C$, and length-scale parameter $\theta_C$.
Following NNGP, we use the kriging variance for $\sigma^2(N,\bv)$:
\begin{equation}\label{eq:variance-GP-approach}
    \sigma^2(N,\bv) = C_{\bv|N} = C(\bv,\bv)-\bC_{\bv,N}\bC_{N}^{-1}\bC_{N,\bv}.
\end{equation}
Consequently, 
\begin{equation*}
    \sigma^2(N(\bv),\bv) = C_{\bv|N(\bv)} = C(\bv,\bv)-\bC_{\bv,N(\bv)}\bC_{N(\bv)}^{-1}\bC_{N(\bv),\bv}.
\end{equation*}

For the conditional mean function $f(\bw,N,\bv)$, we propose an additive structure that superimposes a non-linear Bayesian non-parametric residual term onto a linear NNGP baseline:
\begin{equation}\label{eq:mean-GP-approach}
    f(\bw, N,\bv) = \underbrace{\bB_{\bv,N}\bw\vphantom{\left(\Lambda^{1/2}\right)}}_{\text{linear term}} + \underbrace{ \tau(\bv,N)\cdot g\left(\Lambda^{1/2}(\bv,N)\bw\right)}_{\text{non-linear term}}.
\end{equation}
Evaluating this at neighbor sets yields:
\begin{equation}\label{eq:mean-GP-approach-2}
    f(\bw_{N(\bv)}, N(\bv),\bv) = \bB_{\bv,N(\bv)}\bw_{N(\bv)} + \tau(\bv,N(\bv))\cdot g\left(\Lambda^{1/2}(\bv,N(\bv))\bw_{N(\bv)}\right).
\end{equation}
Here, $g\sim \mathcal{GP}(0, R)$ is an isotropic GP on $\bbR^m$ with a Mat\'ern correlation function $R(\|\cdot\|)$ with smoothness $3/2$.

To intuitively understand the non-linear component, we detail the formulations of $\tau$ and $\Lambda$ below.
The overall magnitude of the non-linearity is governed by an amplitude function $\tau(\bv, N)$, which decays with the minimum spatial distance $l = d(\bv,N)$ between the query location and its neighbor set:
\begin{equation}\label{eq:tau}
    \tau^2(\bv,N) = \me^{\theta_{\tau,1}}l^{\theta_{\tau,2}},
\end{equation}
where $\theta_{\tau,1}\in\bbR$ and $\theta_{\tau,2}>0$ are model parameters.
This formulation automatically enforces the continuity property:
as $\bv \goto \bu\in N$, $l\downarrow 0$, causing $\tau\downarrow 0$, which forces
$f(\bw,N,\bv)\goto\bB_{\bu,N}\bw = w(\bu)$.
Moreover, consider a maximum-minimum-distance (maximin) ordering scheme, where the reference locations are selected sequentially to maximize their distance from previously ordered points. 
Denote $l_i = d(\bs_i, N(\bs_i))$.
It follows that 
$$
l_i = d(\bs_i, \cS_i)\geq d(\bs_{i+1}, \cS_i) \geq d(\bs_{i+1}, \cS_i\cup \{\bs_i\}) = d(\bs_{i+1}, \cS_{i+1}) = l_{i+1},
$$
which indicates that $l_i$ is decreasing in $i$.
As a result, the conditioning locations $\cS_i$ become denser and denser, 
and the non-linear amplitude $\tau_i^2 = \tau^2(\bs_i,N(\bs_i))$ decreases, 
implying that the conditional distribution of $\bw_{\cS\backslash\cS_i}$ given $\bw_{\cS_i}$ becomes increasingly Gaussian.
\textcite{katzfuss2024scalable} argues that such a behavior generally holds for stochastic processes with quasi-quadratic log-likelihoods.

To model the screening effect, motivated by the role of length-scale parameters in anisotropic GPs, we introduce a weight matrix $\Lambda$ to rescale $\bw$ entry-wise in the non-linear part. 
For $N=\{\bv_1,\ldots,\bv_m\}$, we take $\Lambda(\bv, N)$ to be a diagonal matrix with the $i$-th diagonal element
\begin{equation*}
    \Lambda_{ii} = \me^{\theta_{\lambda,1}}\me^{\theta_{\lambda,2} d(\bv,\bv_i)}.
\end{equation*}
Here, $\theta_{\lambda,1}\in\bbR$ and $\theta_{\lambda,2}<0$ are model parameters.
The parametrization implies that points in the conditioning set that are closer to $\bv$ have a higher impact than points away from $\bv$.

Regarding the invariance properties, it can be verified that any stationary C guarantees that the conditional variance (\ref{eq:variance-GP-approach}) and the conditional mean (\ref{eq:mean-GP-approach}) are both translation invariant.
However, $f$ is generally not permutation invariant because $g$ is not a set function.
Nonetheless, given $\bv$, we can treat $\bar f_N(\bw) = f(\bw, N, \bv)$ as a GP over $\bw$, with mean $\bB_{\bv,N}\bw$ and covariance 
\begin{equation*}
    \begin{split}
        \text{Cov}(\bar f_N(\bw), \bar f_N(\bw')) = \tau^2(\bv,N)\cdot R\left(\sqrt{\sum_{i=1}^m \Lambda_{ii}(w_i-w_i')^2}\right),
    \end{split}
\end{equation*}
which indicates that the prior distribution of $\bar f_N$ is invariant to simultaneous permutations of neighbor locations $N$ and their associated values $\bw$.
% As a result, the probability distribution of $\bar f_N$ is invariant if we make the same reordering to $N$ and the argument $\bw$.
Hence, the conditional mean function $f$ in (\ref{eq:mean-GP-approach}) can be viewed as being \textit{probabilistically} permutation invariant.

% \begin{rem}
%     We may alternatively follow BaTraMaSpa to set $\sigma^2(N(\bv),\bv) = (\alpha-1)\me^{\theta_{\sigma,1}}l_i^{\theta_{\sigma,2}}\bar \sigma^2$ with $\bar\sigma^2\sim \text{IG}(\alpha,1)$. In fact, the inverse polynomial decaying of $\sigma^2(N(\bv),\bv)$ in terms of $l_i$ is motivated by the behavior of Gaussian processes with Mat\'ern covariance. So, there should be no essential difference.
% \end{rem}

\subsection{Probability Distributions of the NNnGP}
Since we assign a GP prior to $g$, the conditional mean function $f$ in (\ref{eq:mean-GP-approach}) becomes random. 
For clarification, $p(\bw_\cS)$ and $p(\bw_\cU|\bw_\cS)$ described in Section~\ref{sec:NNnGP-general} should be viewed as \textit{conditioned} on $f$.
In this section, we re-derive $p(\bw_\cS)$ and $p(\bw_\cU|\bw_\cS)$ after marginalizing out $g$. 
All proofs are deferred to Supplement Section~\ref{sec:proofs}. 

We first introduce a few notations.
For $i=1,\ldots,k$, let $w_i = w(\bs_i)$;
denote $h_i = \bB_{\bs_i, N(\bs_i)}\bw_{N(\bs_i)}$ and $F_i = C_{\bs_i|N(\bs_i)}$ as the NNGP's conditional mean and variance for $w_i$ given $\bw_{1:(i-1)}$, respectively.
For the first $m$ variables $\bw_{1:m}=(w_1,\ldots,w_m)^\top$, we simply assume that $\bw_{1:m}\sim N(\bzero, \bC_{mm})$, where $\bC_{mm} = C(\bs_{1:m},\bs_{1:m})$. 
For the remaining reference locations $\cS' = \{\bs_{m+1},\ldots,\bs_k\}$, 
define $\tau_i = \tau(\bs_i,N(\bs_i))$,  
and $\tilde\bw_i=\Lambda^{1/2}_{\bs_i,N(\bs_i)}\bw_{N(\bs_i)}\in\bbR^m$ as the $i$-th rescaled neighbor realizations.
Aggregating these across $\cS'$ yields
$\bw_{\cS'} = (w_{m+1},\ldots,w_k)^\top$, 
$\bh_{\cS'} = (h_{m+1},\ldots,h_k)^\top$, 
$\bD_{\btau_{\cS'}} = \mathrm{diag}(\tau_{m+1},\ldots,\tau_k)$,
$\bD_{\bF_{\cS'}} = \mathrm{diag}(F_{m+1},\ldots,F_k)$,
and $\tilde\bw_{\cS'}=\{\tilde\bw_{m+1},\ldots,\tilde\bw_k\}$.

Regarding the joint density $p(\bw_\cS)$, we have the following result.
\begin{pro}\label{pro:density-ws}
    Under the conditional variance function (\ref{eq:variance-GP-approach}) and the conditional mean function (\ref{eq:mean-GP-approach}), the marginal density $p(\bw_\cS)$ admits the closed form
    \begin{equation}\label{eq:density-ws}
        p(\bw_\cS) = p_{\mathrm{NNGP}}(\bw_\cS)\sqrt{\frac{|\bG|}{|\bR_{\tilde\bw_{\cS'}}|}}\mathrm{exp}\left\{\frac{1}{2} \ba^\top\bG\ba\right\},
    \end{equation}
    where $p_{\mathrm{NNGP}}(\bw_\cS) = \prod_{i=1}^k N(w_i|h_i,F_i)$ is the density of $\bw_\cS$ under the NNGP model; $\bG^{-1} = \bD_{\btau_{\cS'}}\bD_{\bF_{\cS'}}^{-1}\bD_{\btau_{\cS'}} + \bR_{\tilde\bw_{\cS'}}^{-1}$, $\ba = \bD_{\btau_{\cS'}}\bD_{\bF_{\cS'}}^{-1}(\bw_{\cS'} - \bh_{\cS'})$,
    and $\bR_{\tilde\bw_{\cS'}} = R(\tilde\bw_{\cS'},\tilde\bw_{\cS'})$ is the covariance matrix of Gaussian process $g$ at inputs $\tilde\bw_{\cS'}$.
\end{pro}
Proposition~\ref{pro:density-ws} characterizes the relationship between the NNnGP and NNGP models. 
Compared with the NNGP density, (\ref{eq:density-ws}) involves an additional factor, which stems from the non-linearity of $f$ and makes $p(\bw_\cS)$ generally non-Gaussian.
As a special case, if $\theta_{\tau,1}\goto -\infty$ in (\ref{eq:tau}), then $\tau(\bv,N)\goto 0$ globally and the non-linear term in $f$ vanishes, indicating that NNnGP should reduce back to NNGP. 
Indeed, it can be verified that if $\theta_{\tau,1}\goto -\infty$, then $|\bG|\goto |\bR_{\tilde\bw_{\cS'}}|$, and $\ba\goto \bzero$, so that $p(\bw_\cS)$ in (\ref{eq:density-ws}) becomes identical to $p_\mathrm{NNGP}(\bw_\cS)$.

\begin{rem}
    The NNGP density $p_{\mathrm{NNGP}}(\bw_\cS)$ can be rewritten as $N(\bw_\cS|\bzero, \tilde\bC)$, where the precision matrix $\tilde\bC^{-1}$ has a sparse reverse Cholesky factor with at most $k(m+1)$ nonzero entries. 
    More specifically, let $\bT$ be the reverse Cholesky factor of $\tilde\bC^{-1}$ such that $\bT$ is lower triangular and $\tilde\bC^{-1} = \bT^\top\bT$. 
    Then, the $(i,j)$-th entry of $\bT$ is given by
    $$
    T_{ij} = \begin{cases}
    1/\sqrt{C_{\bs_i|N(\bs_i)}}, & \text{if } j = i;\\
    -T_{ii}[\bB_{\bs_i,N(\bs_i)}]_{\# (\bs_j, N(\bs_i))}, & \text{if } \bs_j \in N(\bs_i);\\
    0, & \text{otherwise},
    \end{cases}
    $$
    where $\# (\bs_j, N(\bs_i))$ denotes the index of $\bs_j$ in $N(\bs_i)$.
    See Proposition 1 in \textcite{katzfuss2021general} for more details.
\end{rem}

Next, consider $\cU=\{\bu_1,\ldots,\bu_r\}\subset\cD\backslash\cS$ as $r$ non-reference locations.
For $i=1,\ldots,r$, similarly as before, denote $h_{\bu_i} = \bB_{\bu_i, N(\bu_i)}\bw_{N(\bu_i)}$,
$F_{\bu_i} = C_{\bu_i|N(\bu_i)}$,
$\tau_{\bu_i} = \tau(\bu_i,N(\bu_i))$,
and $\tilde\bw_{\bu_i} = \Lambda^{1/2}(\bu_i,N(\bu_i))\bw_{N(\bu_i)}$,
which aggregate to $\bh_\cU=(h_{\bu_1},\ldots,h_{\bu_r})^\top$,
$\bD_{\bF_\cU} = \mathrm{diag}(F_{\bu_1},\ldots,F_{\bu_r})$,
$\bD_{\btau_\cU} = \mathrm{diag}(\tau_{\bu_1},\ldots, \tau_{\bu_r})$,
and $\tilde\bw_\cU = \{\tilde\bw_{\bu_1},\ldots,\tilde\bw_{\bu_r}\}$.
Regarding $p(\bw_\cU|\bw_\cS)$, we have the following result.
\begin{pro}\label{pro:density-wu-given-ws}
    Let $\cU=\{\bu_1,\ldots,\bu_r\}\subset\cD\backslash\cS$. 
    Under the conditional variance function (\ref{eq:variance-GP-approach}) and the conditional mean function (\ref{eq:mean-GP-approach}), $p(\bw_\cU|\bw_\cS)$ is jointly Gaussian
    \begin{equation}\label{eq:density-wu-given-ws}
        \begin{split}
        p(\bw_\cU|\bw_\cS) = N(\bw_\cU|&\bh_\cU + \bD_{\btau_\cU}\widetilde\bB_{\tilde\bw_\cU,\tilde\bw_{\cS'}}\bG\ba, \\
        &\bD_{\bF_\cU} + \bD_{\btau_\cU}(\bR_{\tilde\bw_\cU|\tilde\bw_{\cS'}} + \widetilde\bB_{\tilde\bw_\cU,\tilde\bw_{\cS'}}\bG\widetilde\bB_{\tilde\bw_{\cS'},\tilde\bw_\cU})\bD_{\btau_\cU}).
        \end{split}
    \end{equation}
    Here, $\widetilde\bB_{\tilde\bw_\cU,\tilde\bw_{\cS'}} = \bR_{\tilde\bw_\cU\tilde\bw_{\cS'}}\bR_{\tilde\bw_{\cS'}}^{-1}$ and $\bR_{\tilde\bw_\cU|\tilde\bw_{\cS'}} = \bR_{\tilde\bw_\cU} - \bR_{\tilde\bw_\cU\tilde\bw_{\cS'}}\bR_{\tilde\bw_{\cS'}}^{-1} \bR_{\tilde\bw_{\cS'}\tilde\bw_\cU}$ are the kriging coefficients and covariance matrices for the Gaussian process $g$ at inputs $\tilde\bw_\cU$ and $\tilde\bw_{\cS'}$, respectively; $\bR_{\tilde\bw_\cU\tilde\bw_{\cS'}} = R(\tilde\bw_\cU,\tilde\bw_{\cS'})$ and $\bR_{\tilde\bw_\cU} = R(\tilde\bw_\cU,\tilde\bw_\cU)$ are defined similarly as $\bR_{\tilde\bw_{\cS'}}$.
\end{pro}

\begin{rem}
    Proposition~\ref{pro:density-wu-given-ws} indicates that the conditioning process $w(\bu)|\bw_\cS$ is a GP on $\cD\backslash\cS$, with mean function
    \begin{equation}\label{eq:wu-given-ws-mean}
        \mu(\bu) = h_\bu + \tau_\bu\widetilde\bB_{\tilde\bw_\bu,\tilde\bw_{\cS'}}\bG\ba, 
    \end{equation}
    and covariance function
    \begin{equation}\label{eq:wu-given-ws-covariance}
        \begin{split}
            \Xi(\bu, \bu') = \delta_{\bu=\bu'} F_\bu + \tau_\bu\tau_{\bu'}(R_{\tilde\bw_\bu\tilde\bw_{\bu'}} - \bR_{\tilde\bw_\bu\tilde\bw_{\cS'}}\bR_{\tilde\bw_{\cS'}}^{-1}\bR_{\tilde\bw_{\cS'}\tilde\bw_{\bu'}}+\widetilde\bB_{\tilde\bw_\bu,\tilde\bw_{\cS'}}\bG\widetilde\bB_{\tilde\bw_{\cS'},\tilde\bw_{\bu'} }).
        \end{split}
    \end{equation}
    We notice that compared with (\ref{eq:vecchia-non-reference}), $w(\bu)|\bw_\cS$ is generally no longer location-wise independent, due to the random $g$ being integrated out.
    Again as a special case, if $\theta_{\tau,1}\goto -\infty$, then $\tau_{\bu, N(\bu)}\goto 0$ globally, which implies that $\mu(\bu)\goto h_\bu$ and $\Xi(\bu, \bu') \goto \delta_{\bu=\bu'} F_\bu$, reducing to the NNGP conditioning process (\ref{eq:wu-given-ws-NNGP}).
\end{rem}

Although in closed forms, $p(\bw_\cS)$ in (\ref{eq:density-ws}) and $p(\bw_\cU|\bw_\cS)$ in (\ref{eq:density-wu-given-ws}) are both computationally intractable when $k$ is large due to the inversion and determinant evaluation of $\bR_{\tilde\bw_{\cS'}}$, which costs $O(k^3)$.
To accelerate inference, we apply the Fully Independent Conditional (FIC) approximation to the kernel $R$ \parencite{quinonero2005unifying}. 
To be more specific, let $\bZ=\{\bz_1,\ldots,\bz_{\tilde m}\}\in\bbR^m$ be $\tilde m$ inducing points. Denote $Q$ as the Nystr\"om approximation of $R$, i.e., $Q(\bx,\bx') = \bR_{\bx,\bZ}\bR^{-1}_{\bZ}\bR_{\bZ,\bx'}$, where $\bR_{\bx,\bZ}=R(\bx,\bZ)$ and $\bR_{\bZ} = R(\bZ,\bZ)$. 
The FIC approximation of $R$ is
\begin{equation}\label{eq:FIC}
    \tilde R(\bx,\bx') = Q(\bx,\bx') + \delta_{\bx=\bx'}(R(\bx,\bx') - Q(\bx,\bx')).
\end{equation}
After replacing $R$ by $\tilde R$, the costs of evaluating $p(\bw_\cS)$ and sampling from $p(\bw_\cU|\bw_\cS)$ can be reduced to $O(k(\tilde m^2+m^3))$ and $O(k(\tilde m r + \tilde m^2 + r^2 + m^3)+r^3)$, respectively. 
See Supplement Section~\ref{sec:low-rank} for more details.

\subsection{A Spatial Regression Model and Inference}
We consider the application of NNnGP in a regression model:
\begin{equation}\label{eq:regression-model}
    y(\bt) = \bx(\bt)^\top\bbeta + w(\bt) + \epsilon(\bt),
\end{equation}
where $\bx(\bt)\in\bbR^{d_x}$ are $d_x$ spatial covariates observed at location $\bt\in\cD$; $w$ is the NNnGP with conditional variance and mean specified in (\ref{eq:variance-GP-approach}) and (\ref{eq:mean-GP-approach}), respectively;
$\epsilon(\bt)\mathop\sim^\text{i.i.d.} N(0,\sigma_\epsilon^2)$ is a Gaussian white-noise process capturing the measurement error or small-scale variations.
The regression model (\ref{eq:regression-model}) can be naturally extended to non-real-valued outcomes using generalized linear models. 
Let $\cT = \{\bt_1,\ldots,\bt_n\}$ be the set of (disjoint) locations where the outcomes $\by_\cT = (y(\bt_1),\ldots, y(\bt_n))^\top$ have been observed. 

The inference procedure depends on whether $\cT\subseteq\cS$ or $\cT\nsubseteq\cS$, and the latter case turns out to be more computationally challenging than the former case.
Nonetheless, in practice, we can always choose a reference set to contain the observed locations.
Therefore, we assume $\cT\subseteq\cS$ here, and leave the alternative case to Supplement Section~\ref{sec:prediction-T-notsubsetof-S}.
Let $\bX_{\cT} = (\bx(\bt_1),\ldots,\bx(\bt_n))^\top$ and $\bw_\cT=(w(\bt_1),\ldots,w(\bt_n))^\top$.
The Bayesian hierarchical model is
\begin{equation}\label{eq:hierarchical-model-subset}
    \begin{split}
        p(\by_\cT|\bw_\cS) &= N(\by_\cT|\bX_\cT\bbeta + \bw_\cT, \sigma^2_\epsilon\bI), \\
        p(\bw_\cS) &= p_{\mathrm{NNGP}}(\bw_\cS)\sqrt{\frac{|\bG|}{|\bR_{\tilde\bw_{\cS'}}|}}\mathrm{exp}\left\{\frac{1}{2} \ba^\top\bG\ba\right\}.
    \end{split}
\end{equation}
The model involves unknown parameters $(\btheta,\bbeta, \sigma^2_\epsilon)$ and latent variables $\bw_\cS$, where $\btheta = \{\sigma^2_C,\allowbreak \theta_C,\allowbreak \theta_{\tau,1},\allowbreak \theta_{\tau,2},\allowbreak \theta_{\lambda,1},\allowbreak \theta_{\lambda,2}\}$ are the NNnGP parameters. 
The joint likelihood can be efficiently evaluated in $O(nd_x + k(\tilde m^2+m^3))$.

Our inferential goal is twofold: estimating the unknown parameters $(\btheta,\bbeta, \sigma^2_\epsilon)$ via empirical Bayes, and obtaining posterior samples for $\bw_\cS$.
VI handles both tasks simultaneously by maximizing the evidence lower bound (ELBO) with respect to the variational parameters and the unknown model parameters.
Crucially, standard VI employs Gaussian variational families, which are inadequate to approximate the complex posterior landscape of $\bw_\cS$ (as will be demonstrated in Section~\ref{sec:illustrations}).
To address this, we employ normalizing flow-based VI.
A normalizing flow constructs a highly expressive, non-Gaussian probability distribution by applying a sequence of invertible and differentiable transformations (i.e., diffeomorphisms) to a simple base distribution.
Specifically, our numerical implementation adopts a Masked Autoregressive Flow \parencite[MAF,][]{papamakarios2017masked}, which leverages masked neural networks to construct diffeomorphisms.
Detailed mathematical formulations of the ELBO and the MAF architecture are provided in Supplement Section~\ref{sec:flow-vi}.

Once the model is trained, making predictions at unobserved locations with uncertainty quantification becomes straightforward. 
Let $\by_{\tilde\cV}$ be the outcomes at $r$ locations $\tilde\cV = \{\bv_1,\ldots,\bv_r\}$, which can be sampled via Algorithm~\ref{alg:prediction}.

\begin{minipage}{0.9\linewidth}
\begin{algorithm}[H]
\caption{Making predictions (when $\cT\subseteq\cS$)}\label{alg:prediction}
%\footnotesize
\begin{algorithmic}
\STATE {\textbf{Input}} Model parameters $(\btheta,\bbeta, \sigma^2_\epsilon)$.
\REPEAT
\STATE Sample $\bw_\cS\sim p(\bw_\cS|\by_\cT)$.
\STATE For locations $\cU_0 = \tilde\cV\cap\cS$, extract $\bw_{\cU_0}$ from $\bw_\cS$.
\STATE For locations $\cU = \tilde\cV\backslash\cS$, sample $\bw_\cU$ from $p(\bw_\cU|\bw_\cS)$ in (\ref{eq:density-wu-given-ws}).
\STATE Combine $\bw_{\tilde\cV} = (\bw_{\cU_0}, \bw_\cU)$.
\STATE Sample $\by_{\tilde\cV}\sim N(\bX_{\tilde\cV}\bbeta + \bw_{\tilde\cV}, \sigma^2_\epsilon\bI)$.
\UNTIL{A desired number of samples are obtained.}
\end{algorithmic}
\end{algorithm}
\end{minipage}
\vspace{.2in}

\subsection{Ordering and the Neighborhood Size}\label{sec:ordering-neighbor}
The decomposition (\ref{eq:autoregressive-model}) and the Vecchia approximation (\ref{eq:vechia-reference}) depend on the ordering of the reference locations.
\textcite{guinness2018permutation} discussed and compared several ordering methods, including sorted coordinate orderings, maximin ordering, middle-out ordering, and completely random orderings.
It has been argued in \textcite{guinness2018permutation} that the maximin ordering can be effective for GP with Mat\'ern covariance in two dimensions, and for dimensions more than two, random orderings can be much better than the sorted coordinate orderings.
Following \textcite{katzfuss2024scalable}, we recommend the maximin ordering, which spaces out the first $i$ locations into a quasi-regular grid with spacing roughly proportional to $O(i^{-1/d})$.

Regarding the neighborhood size $m$, similar to NNGP and NNMP, increasing $m$ allows each location to borrow information from a larger neighbor set, which improves the model fidelity. 
However, it also increases the computational cost.
For NNnGP, the neighborhood size $m$ also determines the input dimension of the Gaussian process $g$ in the conditional mean (\ref{eq:mean-GP-approach}).
A large $m$ implies a GP defined in a high-dimensional space, which, in general, is difficult to learn.
Nonetheless, the low-rank approximation provides regularization to $g$. 
In addition, the NNnGP does not need to fully infer $g$ on the entire domain, since the inputs of $g$ are always the rescaled neighbor values, which are highly structured and correlated. 
As a result, $g$ may only need to be inferred on a low-dimensional manifold.
See Section~\ref{sec:illustrations} for some numerical evidence.
That being said, we recommend a small or medium neighbor size $m$ (e.g., below $20$) to balance the model fidelity and estimation efficiency.

\section{Illustrations}\label{sec:illustrations}

We conduct numerical experiments on two synthetic datasets and a real dataset to evaluate the inference and prediction performance of NNnGP. 
The inference of NNnGP is programmed in Python and leverages open-source libraries 
\texttt{NumPyro}\footnote{\url{https://num.pyro.ai/en/stable/}} for implementing HMC and
VI with mean-field and low-rank Gaussian variational families, and \texttt{FlowJAX}\footnote{\url{https://danielward27.github.io/flowjax/}} for implementing normalizing flow-based VI.
\texttt{NumPyro} is a lightweight NumPy backend for \texttt{Pyro}\footnote{\url{https://pyro.ai/}}, which relies on \texttt{JAX}\footnote{\url{https://github.com/jax-ml/jax}} for automatic differentiation (AD) and Just-In-Time compilation to GPU/CPU.
Compared with \texttt{Pyro}, \texttt{NumPyro} provides a significant speedup for HMC and NUTS.
Unlike \texttt{NumPyro}, \texttt{FlowJAX} is specifically designed for building and working with bijections and normalizing flows, which is also based on \texttt{JAX} for AD.
Code for reproducing the numerical results is available at \url{https://github.com/ffpphh/NNnGP}.

\subsection{Simulation Studies}\label{sec:simulation}
The first experiment is designed to validate the normalizing flow-based VI and demonstrate 
NNnGP's adaptivity to the degree of non-Gaussianity of the data.
We created a regular grid of $100\times 100$ resolution on a square domain $[0,5]^2$, and randomly selected $k=500$ points as the reference points $\cS$ ordered by maximin. 
The data were simulated on the grid according to the spatial regression model (\ref{eq:regression-model}) with the latent field $w$ following the NNnGP structure specified in Section~\ref{sec:NNnGP-GP}.
We set the neighborhood size $m=10$ and drew the non-linear function $g$ from an isotropic GP with a Mat\'ern $3/2$ correlation, approximated by FIC with $\tilde m =50$ pre-specified inducing points.
Further simulation details are provided in Supplement Section~\ref{sec:simulation-model}.

\begin{figure}[!htb]
    \centering
    \begin{subfigure}{0.82\textwidth}
        \centering
        \includegraphics[width=\linewidth]{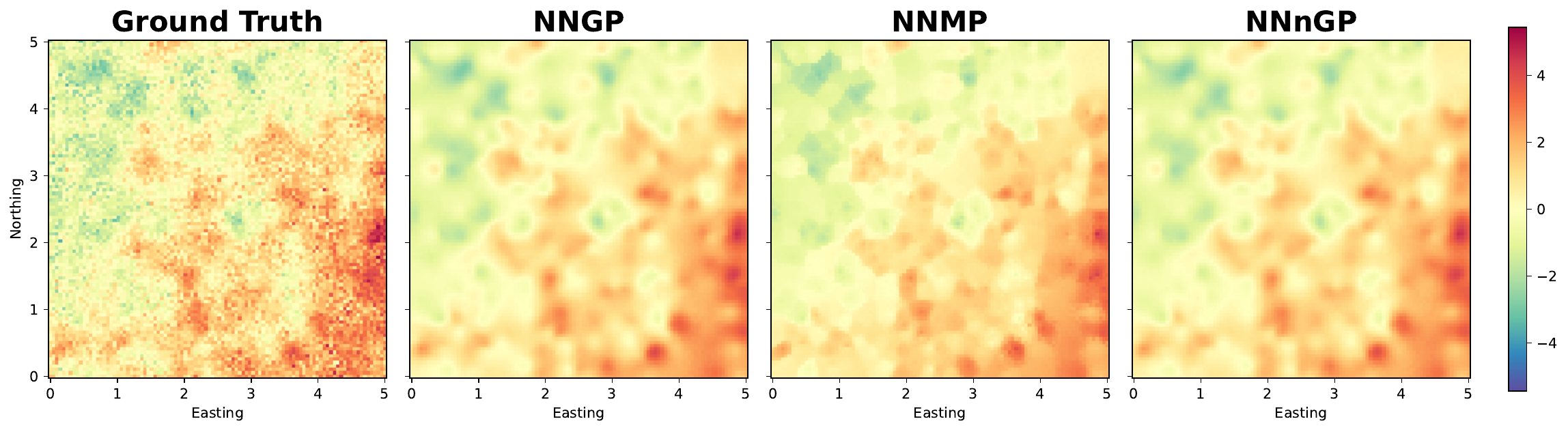}
        \caption{Weak non-linearity.}
    \end{subfigure}\vspace{.05in}\hfill
    
    \begin{subfigure}{0.82\textwidth}
        \centering
        \includegraphics[width=\linewidth]{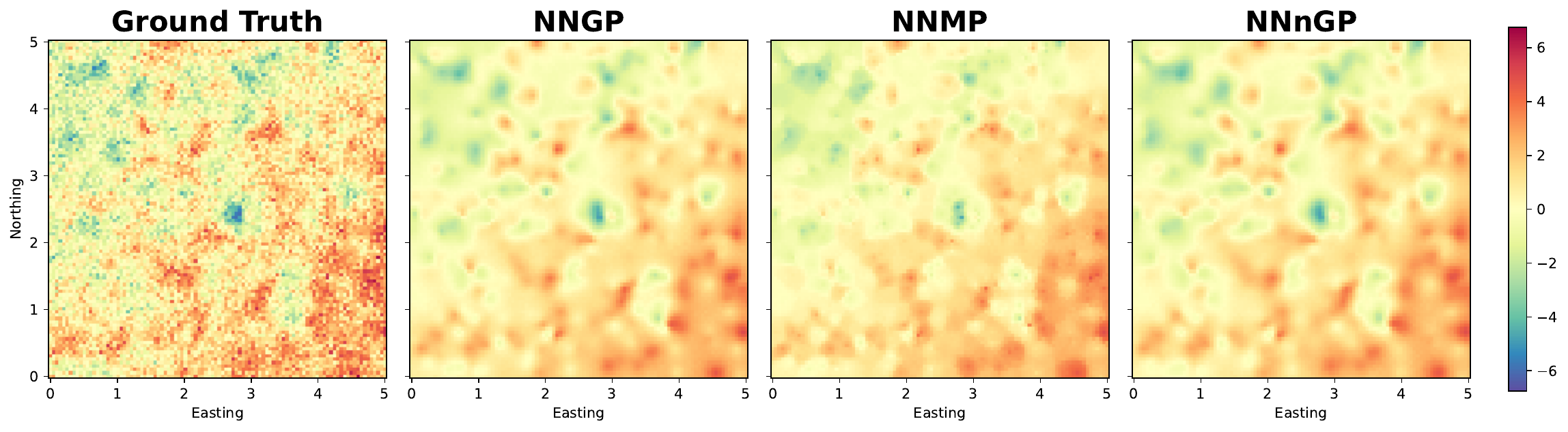}
        \caption{Medium non-linearity.}
    \end{subfigure} \vspace{.05in}\hfill
    
    \begin{subfigure}{0.82\textwidth}
        \centering
        \includegraphics[width=\linewidth]{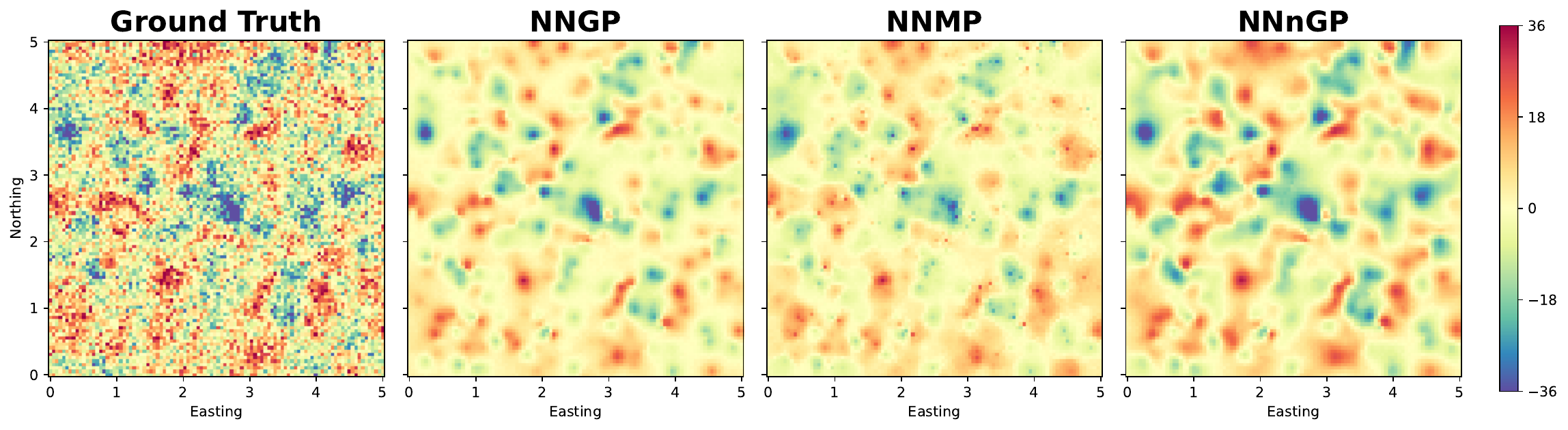}
        \caption{Strong non-linearity.}
        \label{fig:prediction-mean-strong-nonlinearity}
    \end{subfigure} 
    \caption{Ground truth and prediction means of $y$.}
    \label{fig:prediction-mean}
\end{figure}

To assess adaptivity, we let $\theta_{\tau,1}$ vary in $\{-5, 0, 5\}$ to induce small, medium, and large magnitudes of
the non-linear term in the conditional mean (\ref{eq:mean-GP-approach}), respectively.
All other true parameter values are summarized in the first rows of Tables \ref{tab:metrics-VI-methods} and \ref{tab:metrics-other-methods}, and the simulated response fields are visualized in Figure~\ref{fig:prediction-mean}. 
We compare the following methods: 
\begin{itemize}
    \item \textbf{NNGP}: The Nearest-neighbor Gaussian processes described in Section~\ref{sec:related models}, implemented via the R package \texttt{spNNGP}\footnote{\url{https://cran.r-project.org/web/packages/spNNGP/index.html}}.
    \item \textbf{NNMP}: The Nearest-neighbor mixture processes with bivariate Gaussian kernels described in Section~\ref{sec:related models}, implemented via the R package \texttt{nnmp}\footnote{\url{https://github.com/xzheng42/nnmp-examples}}.
    \item \textbf{NNnGP-HMC}: The NNnGP model with the latent $\bw_\cS$ inferred by HMC with other parameters fixed at their true values. In practice, parameters are unknown. Therefore, this result is only for a benchmark.
    \item \textbf{NNnGP-VI-MF, NNnGP-VI-LR, and NNnGP-VI-NF}: The NNnGP model with the latent $\bw_\cS$ inferred by VI with mean-field Gaussian, low-rank Gaussian (rank $50$), and normalizing-flow variational families, respectively. 
    An MAF was implemented for the normalizing flow (see Supplement Section~\ref{sec:flow-vi} for more details). 
    The MAF architecture employs $3$ transformation layers, each using a masked conditioner parameterized by a two-layer neural network ($512$ hidden units, ReLU activation).
\end{itemize}

\begin{table}[tb]
\centering
\scriptsize
\caption{\small Synthetic data analysis: parameter estimates and prediction performance for NNnGP inferred by various VI methods (HMC with true parameters as a benchmark). CIs are $95\%$ ($2.5\%\sim 97.5\%$). }
\label{tab:metrics-VI-methods}
\begin{tabular}{*{11}{c}}
\toprule
& $\sigma_{C}$ & $\theta_{C}$ & $\theta_{\tau,1}$ & $\theta_{\tau,2}$ & $\theta_{\lambda,1}$ & $\theta_{\lambda,2}$ & RSR\% & CRPS & CI cover\% & CI width \\
\midrule
\multicolumn{11}{c}{Weak non-linearity ($\theta_{\tau,1}=-5$)}  \\
Truth & 1.00 & 0.20 & -5.00 & 0.10 & 0.00 & -2.00 & \textemdash & \textemdash & \textemdash & \textemdash \\
NNnGP-HMC & \textemdash & \textemdash & \textemdash & \textemdash & \textemdash & \textemdash & 47.00 & 0.35 & 94.46 & 2.40 \\
NNnGP-VI-MF & 0.55 & 0.12 & -3.13 & 0.02 & 0.25 & -3.94 & 54.96 & 0.42 & 95.66 & 3.06 \\
NNnGP-VI-LR & 0.55 & 0.13 & -3.13 & 0.02 & 0.25 & -3.87 & 54.95 & 0.42 & \textbf{95.79} & 3.07 \\
NNnGP-VI-NF & \textbf{0.91} & \textbf{0.18} & \textbf{-3.87} & \textbf{0.04} & \textbf{-0.13} & \textbf{-1.89} & \textbf{47.34} & \textbf{0.36} & 94.42 & \textbf{2.40} \\
\midrule
\multicolumn{11}{c}{Medium non-linearity ($\theta_{\tau,1}=0$)}  \\
Truth & 1.00 & 0.20 & 0.00 & 0.10 & 0.00 & -2.00 & \textemdash & \textemdash & \textemdash & \textemdash \\
NNnGP-HMC & \textemdash & \textemdash & \textemdash & \textemdash & \textemdash & \textemdash & 61.00 & 0.61 & 94.54 & 4.23 \\
NNnGP-VI-MF & 0.45 & 0.18 & -0.55 & 0.11 & \textbf{-0.54} & -5.30 & 70.51 & 0.71 & 93.09 & 4.55 \\
NNnGP-VI-LR & 0.46 & \textbf{0.18} & \textbf{-0.54} & \textbf{0.11} & -0.56 & -5.47 & 70.43 & 0.70 & 93.16 & 4.57 \\
NNnGP-VI-NF & \textbf{0.71} & 0.17 & 0.60 & 0.26 & -1.24 & \textbf{-1.41} & \textbf{62.37} & \textbf{0.62} & \textbf{95.24} & \textbf{4.40} \\
\midrule
\multicolumn{11}{c}{Strong non-linearity ($\theta_{\tau,1}=5$)}  \\
Truth & 1.00 & 0.20 & 5.00 & 0.10 & 0.00 & -2.00 & \textemdash & \textemdash & \textemdash & \textemdash\\
NNnGP-HMC & \textemdash & \textemdash & \textemdash & \textemdash & \textemdash & \textemdash & 71.00 & 6.13 & 94.93 & 42.60 \\
NNnGP-VI-MF & 0.48 & 0.09 & 2.57 & 0.65 & \textbf{0.00} & \textbf{-2.00} & 96.67 & 10.43 & 26.43 & 9.57 \\
NNnGP-VI-LR & 0.49 & 0.09 & 2.57 & 0.65 & \textbf{0.00} & \textbf{-2.00} & 96.68 & 10.43 & 26.39 & 9.57 \\
NNnGP-VI-NF & \textbf{1.14} & \textbf{0.18} & \textbf{4.96} & \textbf{0.08} & 0.02 & -1.99 & \textbf{71.34} & \textbf{6.12} & \textbf{94.98} & \textbf{42.73} \\
\bottomrule
\end{tabular}
\end{table}

All models were trained using observations on the $500$ reference locations with the same maximin ordering and the same number of neighbors as in the data-generating process. 
The remaining observations were withheld to assess predictive performance.
For MCMC-based methods (NNGP, NNMP, and NNnGP-HMC), we ran $10,000$ iterations with a $5,000$-step burn-in, retaining every $5$th sample to yield $1,000$ posterior draws. 
For VI-based methods (NNnGP-VI-MF, NNnGP-VI-LR, and NNnGP-VI-NF), the flow parameters, model parameters, and the inducing points for the FIC approximation are jointly optimized by maximizing the ELBO using the Adam optimizer for $15,000$ iterations. After training, $1000$ samples were generated from the variational distribution for prediction.
We use RMSE-observations standard deviation ratio \parencite[RSR\footnote{RSR is the ratio of the root mean square error (RMSE) to the standard deviation of the observed data, which is standardized and can be used across different scenarios.},][]{moriasi2007model} to measure the standardized prediction error, and use the continuous ranked probability score \parencite[CRPS,][]{gneiting2007probabilistic} to assess probabilistic forecasts.
Smaller RSR and CRPS indicate better prediction accuracy.
We also report the $95\%$ credible interval (CI) coverage rate and average width to assess uncertainty quantification.

Table~\ref{tab:metrics-VI-methods} details the parameter estimates (of $\btheta$) and performance metrics for NNnGP under various inference schemes. 
Remarkably, NNnGP-VI-NF achieves predictive performance matching the HMC benchmark across all three scenarios, confirming that the MAF architecture effectively captures the true posterior. 
In comparison, NNnGP-VI-MF and NNnGP-VI-LR exhibit severe predictive degradation in the strong non-linearity regime. 
This failure highlights a key motivation for our approach: under strong non-linearity, the latent field $\bw_\cS$ admits a highly complex, non-Gaussian posterior landscape that standard Gaussian variational families cannot accommodate.
We also note a minor identifiability issue between the parameters governing $\tau$ and $\Lambda$, though it does not affect predictive performance.

\begin{table}[tb]
\centering
\scriptsize
\caption{\small Synthetic data analysis: parameter estimates and prediction performance for NNGP, NNMP, and NNnGP-VI-NF. CIs are $95\%$ ($2.5\%\sim 97.5\%$).}
\label{tab:metrics-other-methods}
\begin{tabular}{*{9}{c}}
\toprule
& $\beta_0$ & $\beta_1$ & $\beta_2$ & $\sigma^2_\epsilon$ & RSR\% & CRPS & CI cover\% & CI width \\
\midrule
Truth & 0.50 & 0.50 & -0.50 & 0.04 & \textemdash & \textemdash & \textemdash & \textemdash \\
\midrule
\multicolumn{9}{c}{Weak non-linearity ($\theta_{\tau,1}=-5$)} \\
\midrule
NNGP & 0.27 & \textbf{0.52} & -0.41 & 0.01 & \textbf{47.27} & \textbf{0.36} & 93.37 & \textbf{2.35} \\
NNMP & 0.42 & 0.47 & \textbf{-0.46} & \textbf{0.04} & 50.61 & 0.38 & \textbf{95.92} & 2.79 \\
NNnGP-VI-NF & \textbf{0.44} & \textbf{0.48} & -0.44 & \textbf{0.04} & 47.34 & \textbf{0.36} & 94.42 & 2.40 \\
\midrule
\multicolumn{9}{c}{Medium non-linearity ($\theta_{\tau,1}=0$)} \\
\midrule
NNGP & 0.90 & 0.40 & \textbf{-0.51} & \textbf{0.02} & 62.99 & 0.63 & 93.64 & \textbf{4.33} \\
NNMP & 0.79 & \textbf{0.46} & -0.53 & 0.09 & 66.64 & 0.67 & 94.33 & 4.60 \\
NNnGP-VI-NF & \textbf{0.53} & \textbf{0.46} & -0.45 & 0.09 & \textbf{62.37} & \textbf{0.62} & \textbf{95.24} & 4.40 \\
\midrule
\multicolumn{9}{c}{Strong non-linearity ($\theta_{\tau,1}=5$)} \\
\midrule
NNGP & 3.57 & -0.66 & \textbf{-0.61} & \textbf{0.01} & 75.41 & 6.51 & 91.74 & \textbf{42.02} \\
NNMP & 0.65 & 1.28 & -1.11 & \textbf{0.07} & 79.38 & 6.83 & 91.01 & 43.43 \\
NNnGP-VI-NF & \textbf{0.58} & \textbf{0.36} & \textbf{-0.58} & 0.09 & \textbf{71.34} & \textbf{6.12} & \textbf{94.98} & 42.73 \\
\bottomrule
\end{tabular}
\end{table}

% \begin{figure}[htbp]
%     \centering
%     \includegraphics[width=0.7\textwidth]{figure/wS_prediction_mean_vs_true.pdf}
%     \caption{The true $\bw_\cS$ versus posterior means in the strong non-linearity case.}
%     \label{fig:ws prediction}
% \end{figure}

Table~\ref{tab:metrics-other-methods} and Figure~\ref{fig:prediction-mean} compare NNnGP-VI-NF (abbreviated as NNnGP thereafter) against the NNGP and NNMP models.
NNnGP achieves comparable performance to NNGP in the weak and medium non-linearity cases, and is better than NNGP in the strong non-linearity case, yielding sharper spatial textures (Figure~\ref{fig:prediction-mean-strong-nonlinearity}) and lower prediction errors. 
The results demonstrate that NNnGP is adaptive to the degree of non-linearity in the local dependence.

\begin{figure}[!b]
    \centering
    \begin{subfigure}{0.36\textwidth}
        \centering
        \includegraphics[width=\linewidth]{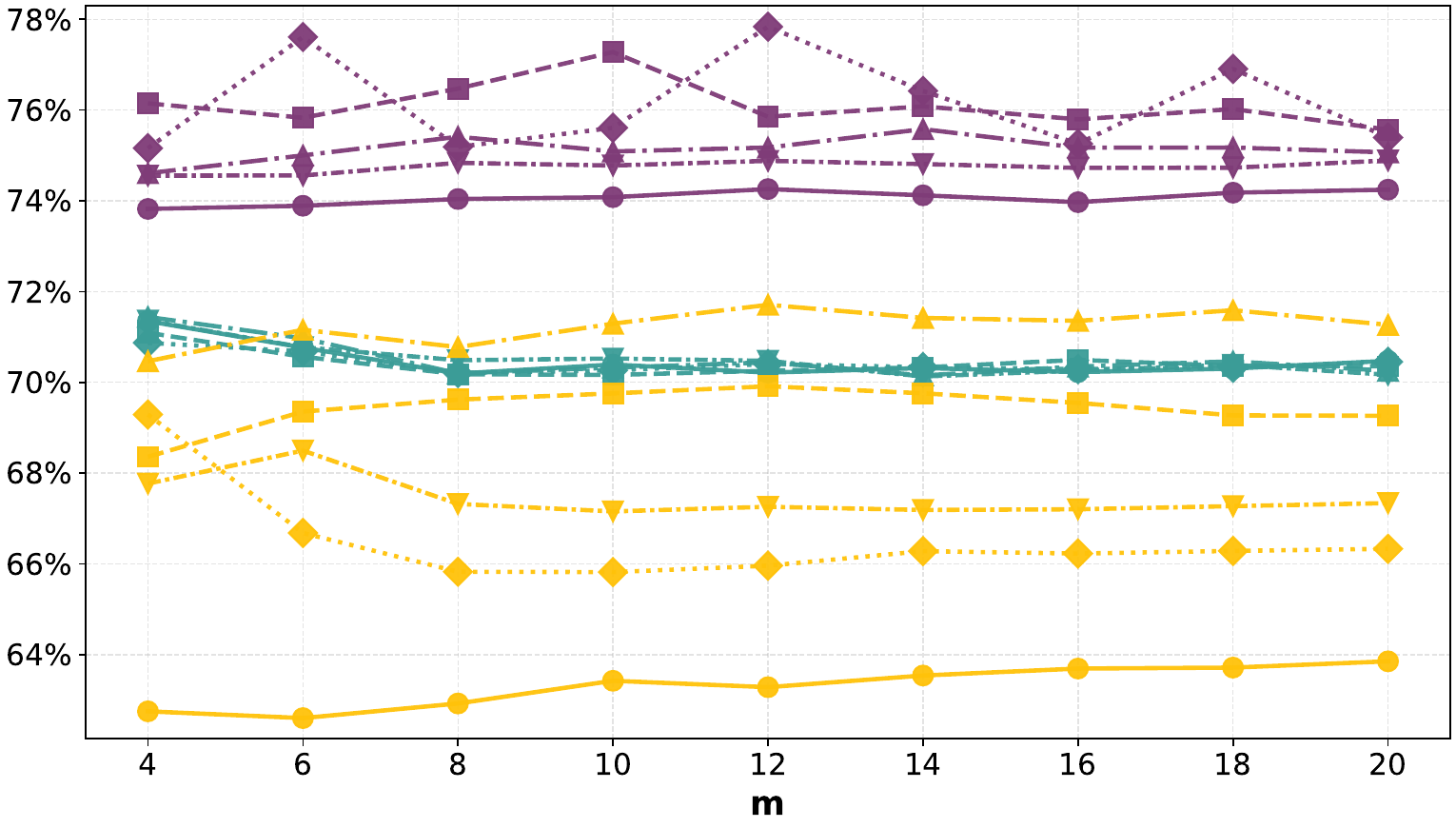}
        \caption{RSR.}
    \end{subfigure} 
    \hspace{.25in}
    \begin{subfigure}{0.36\textwidth}
        \centering
        \includegraphics[width=\linewidth]
        {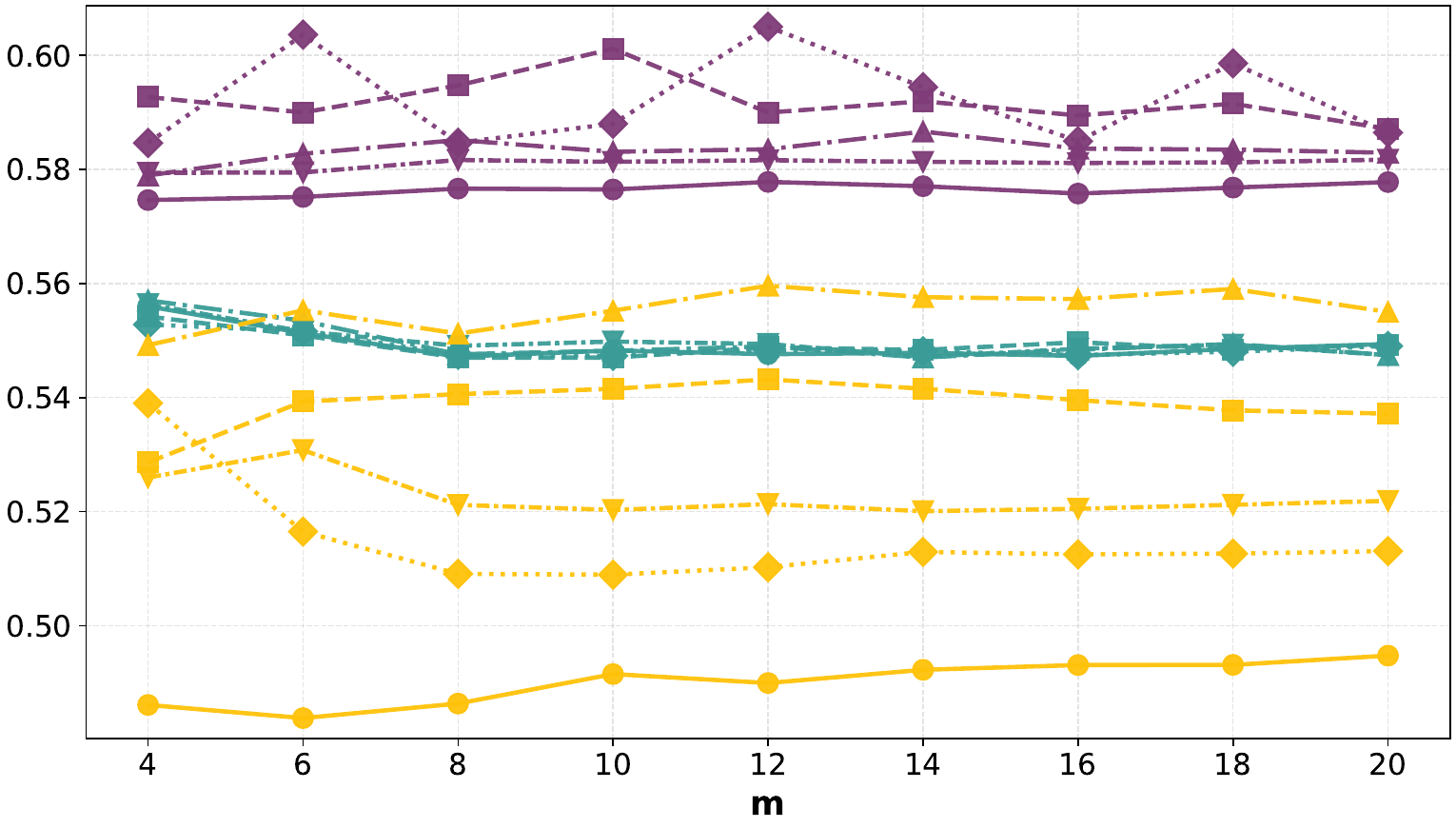}
        \caption{CRPS.}
    \end{subfigure} 

    \vspace{.1in}
    \begin{subfigure}{0.35\textwidth}
        \centering
        \includegraphics[width=\linewidth]{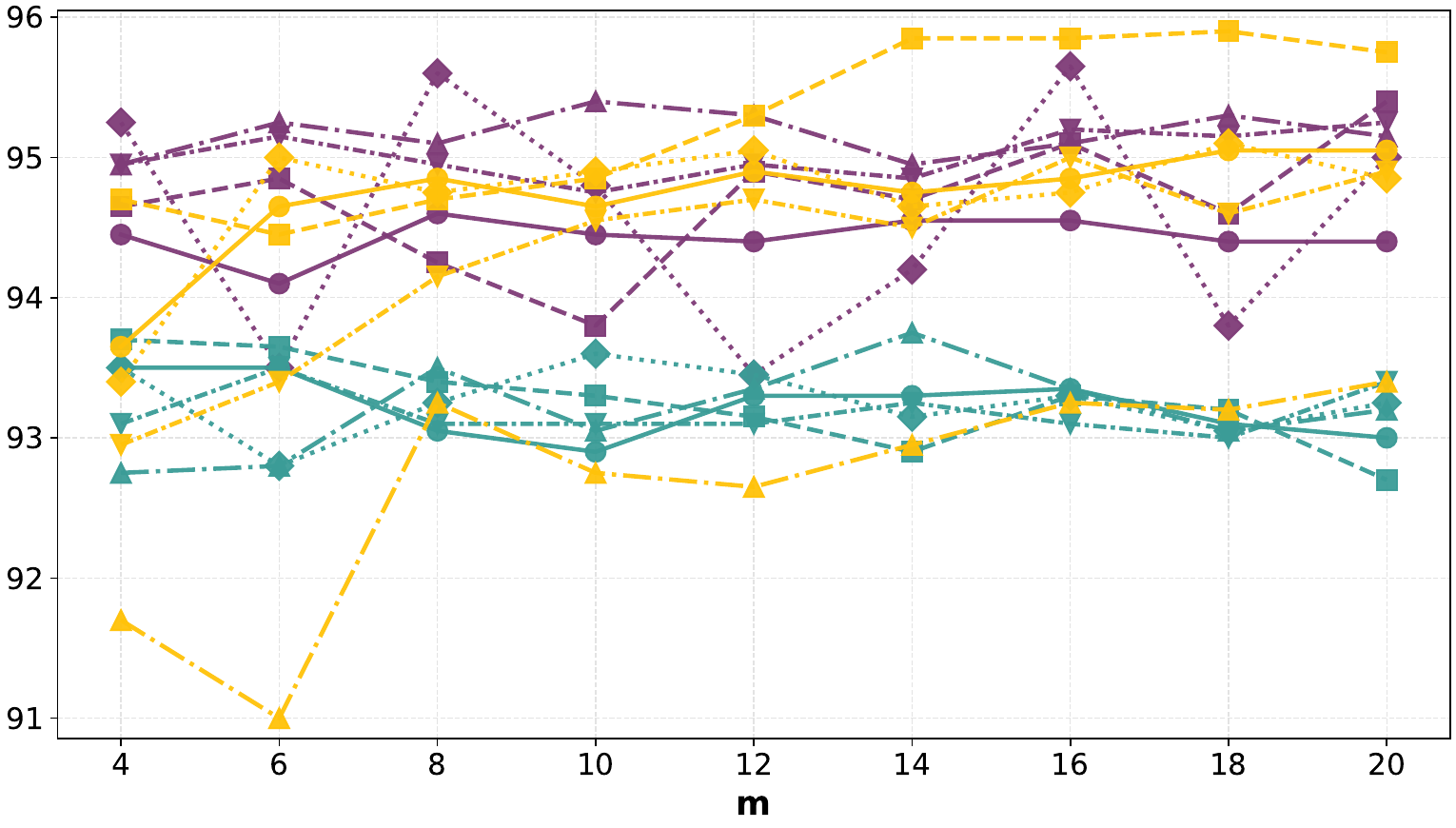}
        \caption{$95\%$ CI coverage rate.}
    \end{subfigure}
    \hspace{.3in}
    \begin{subfigure}{0.35\textwidth}
        \centering
        \includegraphics[width=\linewidth]{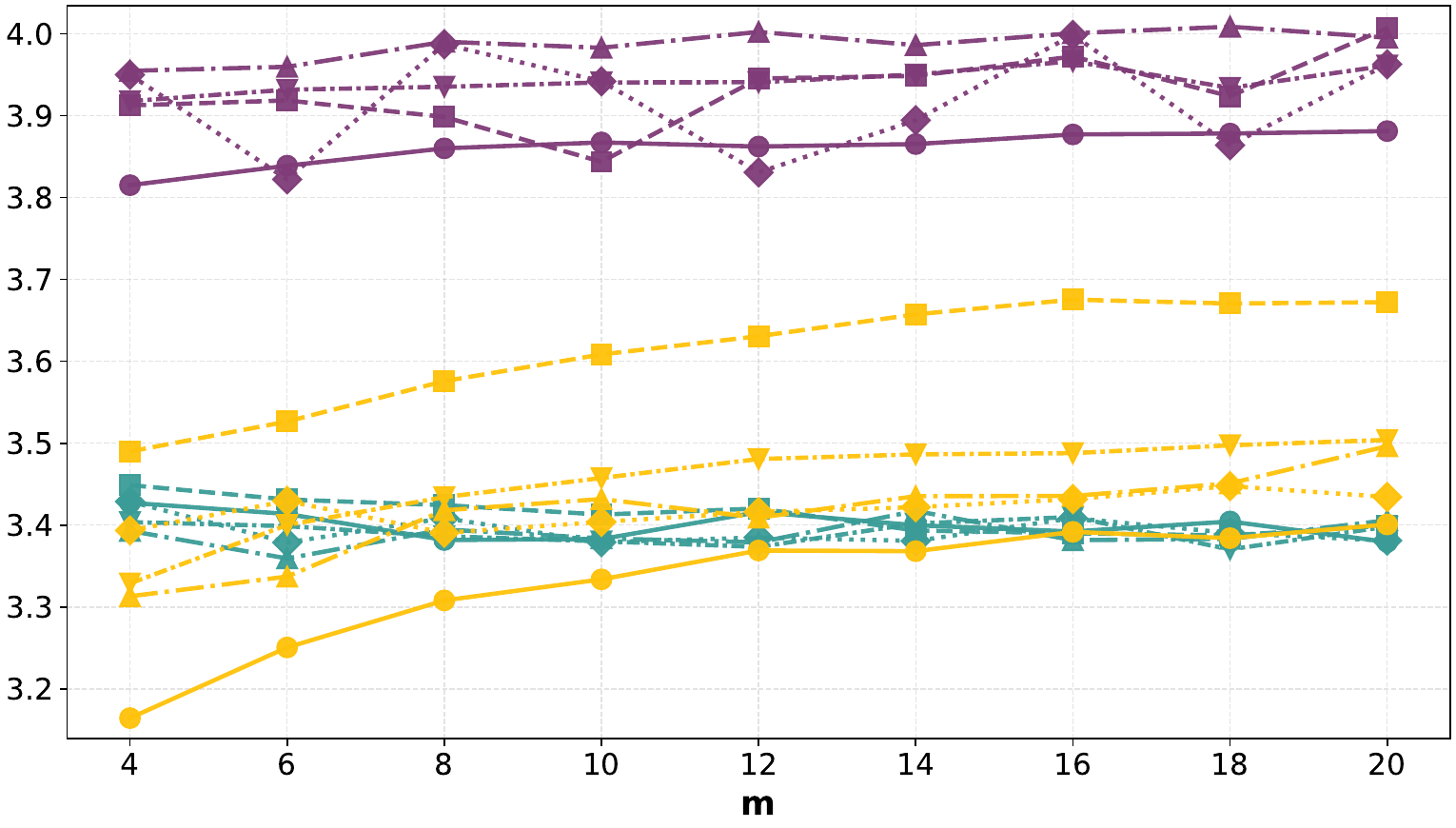}
        \caption{$95\%$ CI width.}
    \end{subfigure}

    \vspace{.1in}
    \includegraphics[height=0.06\textwidth]{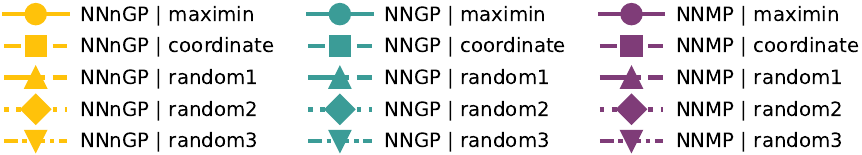}
    \caption{Predictive performance with different orderings and neighbor sizes.}
    \label{fig:result-simu-2}
\end{figure}

The second experiment is designed to test NNnGP under structural misspecification and hyperparameter variations. 
We created a regular grid of size $50\times 50=2500$ on a square domain $[0,5]^2$, and randomly chose $k=500$ points as the reference set, which were ordered by maximin.
We generated the responses according to the NNnGP framework in Section~\ref{sec:NNnGP-general} with a conditional mean function
\begin{equation}\label{eq:mean-simu-2}
    f(\bw_{N(\bv)}, N(\bv),\bv) = \bB_{\bv, N_{10}(\bv)}\bw_{N_{10}(\bv)} + \sin(4\bB_{\bv, N_{2}(\bv)}\bw_{N_{2}(\bv)}),
\end{equation}
and a conditional variance function $\sigma^2(N(\bv),\bv) = C_{\bv|N(\bv)}$.
Here, $N_{j}(\bv)$ denotes the $j$-nearest neighbors, and $C$ is an isotropic Mat\'ern correlation function with smoothness $3/2$.
We note that the true conditional mean is mis-specified by the Bayesian non-parametric formulation (\ref{eq:mean-GP-approach-2}), which assumes a matching number of active neighbors in the linear and non-linear terms.

We compared NNGP, NNMP, and NNnGP across neighborhood sizes $m$ varying from $4$ to $20$ and evaluated five different reference orderings: the correct maximin, sorted coordinate, and three random orderings. 
The predictive performance is shown in Figure~\ref{fig:result-simu-2}.
As expected, NNnGP equipped with the correct maximin ordering yields the best performance. 
Interestingly, even when the ordering is completely randomized---thereby violating the regularized homogeneity condition (\ref{eq:conditional-unified})---NNnGP still delivers predictive accuracy comparable to, or exceeding, both NNGP and NNMP.
Furthermore, increasing the neighborhood size $m$ does not significantly affect NNnGP's performance. 
Together, these results demonstrate the robustness of the NNnGP framework against poorly-tuned hyperparameters, unsuitable reference-location orderings, and structural mis-specifications in the non-linear conditional means. (Similar robustness is observed in additional simulations presented in Supplement Section~\ref{sec:additional-simu}).

\subsection{PRISM Precipitation Data Analysis}\label{sec:precipitation-data}
The study of precipitation anomalies is of paramount importance for extreme weather monitoring and mitigation, water resource management, and understanding climate change.
Extreme precipitation events, such as droughts and floods, often exhibit complex, non-stationary spatial dependencies and sharp regional boundaries that pose significant challenges to traditional Gaussian process models.
In this section, we analyze the total precipitation anomalies in Montana, USA, during October 2025.
The goal is not only to assess global interpolation accuracy but also to examine whether the proposed nonlinear conditional mean structure in NNnGP improves the prediction of tail events.

\begin{figure}[htbp]
    \centering
    \begin{subfigure}[b]{0.49\textwidth}
        \centering
        \includegraphics[width=\textwidth]{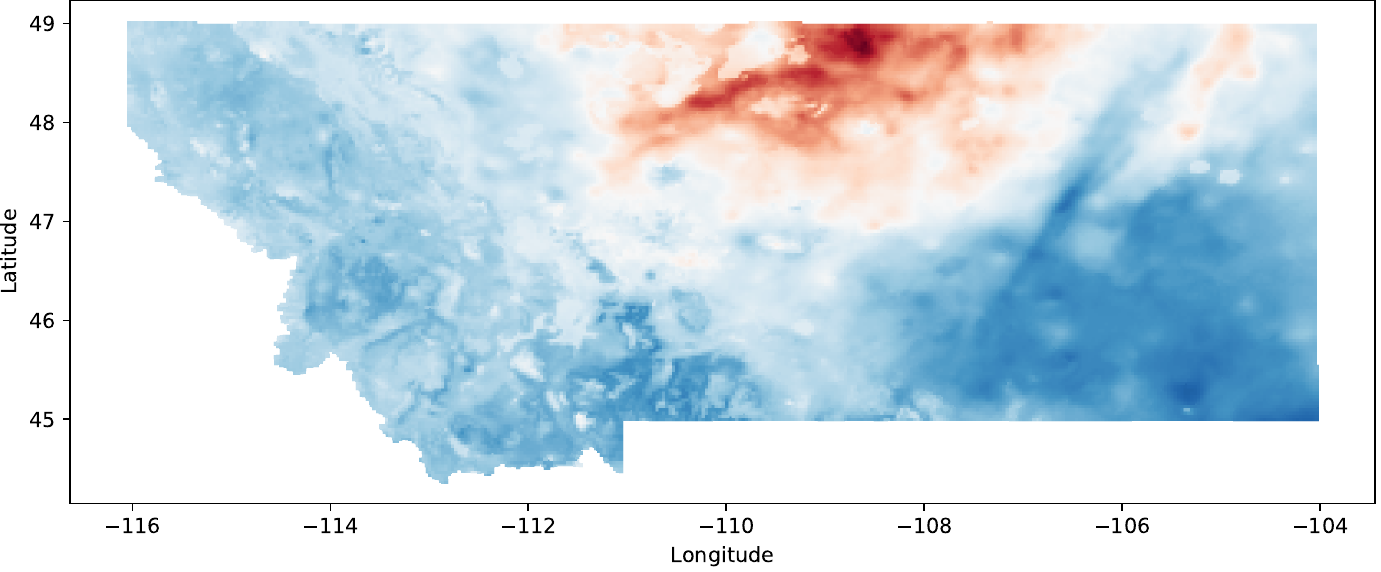}
        \caption{Ground truth}
        \label{fig:precipitation-truth}
    \end{subfigure}\hfill
    \begin{subfigure}[b]{0.49\textwidth}
        \centering
        \includegraphics[width=\textwidth]{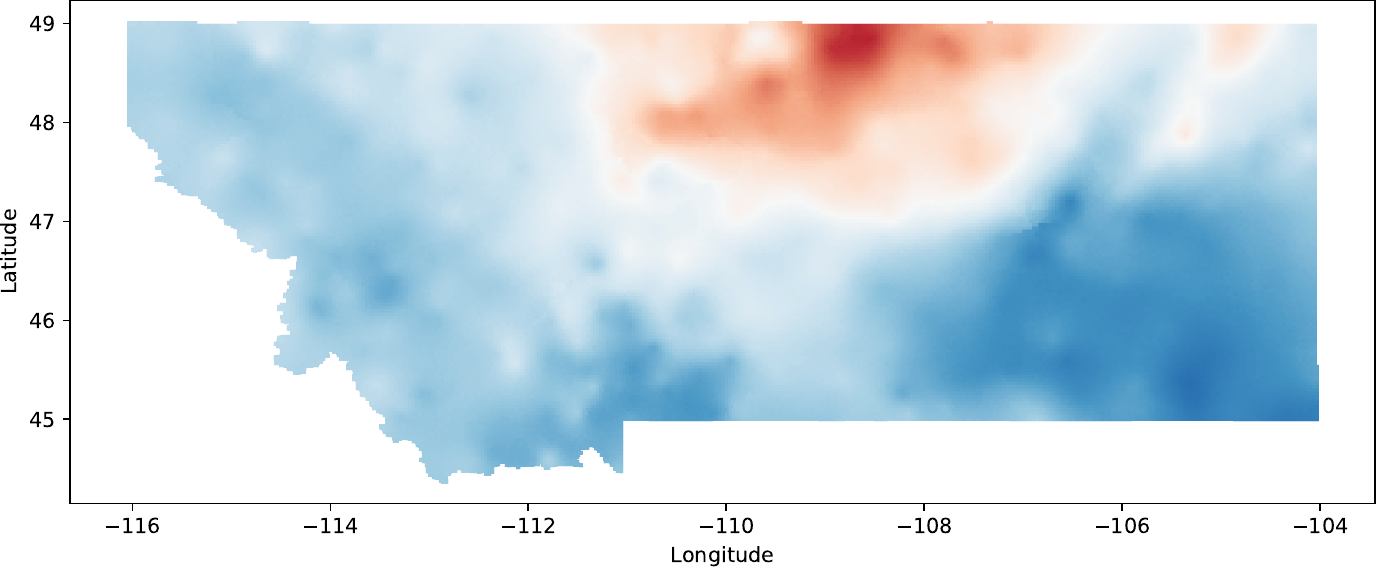}
        \caption{NNGP}
        \label{fig:precipitation-NNGP}
    \end{subfigure}
    \vspace{.1in}\hfill
    
    \begin{subfigure}[b]{0.49\textwidth}
        \centering
        \includegraphics[width=\textwidth]{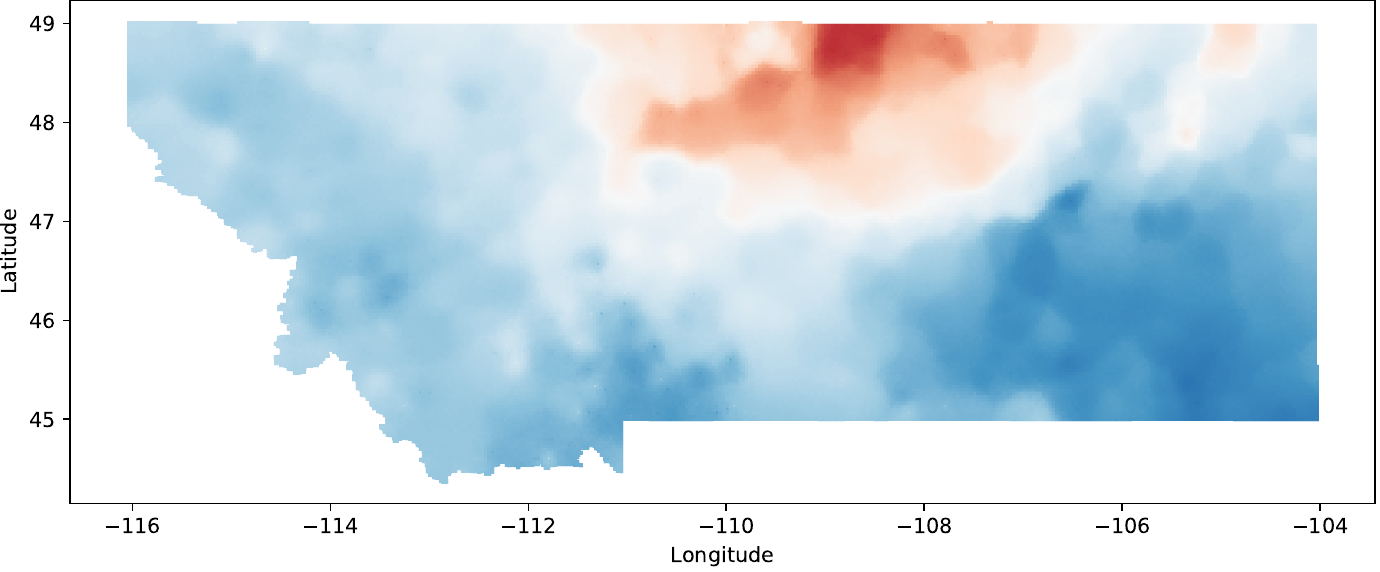}
        \caption{NNMP}
        \label{fig:precipitation-NNMP}
    \end{subfigure}\hfill
    \begin{subfigure}[b]{0.49\textwidth}
        \centering
        \includegraphics[width=\textwidth]{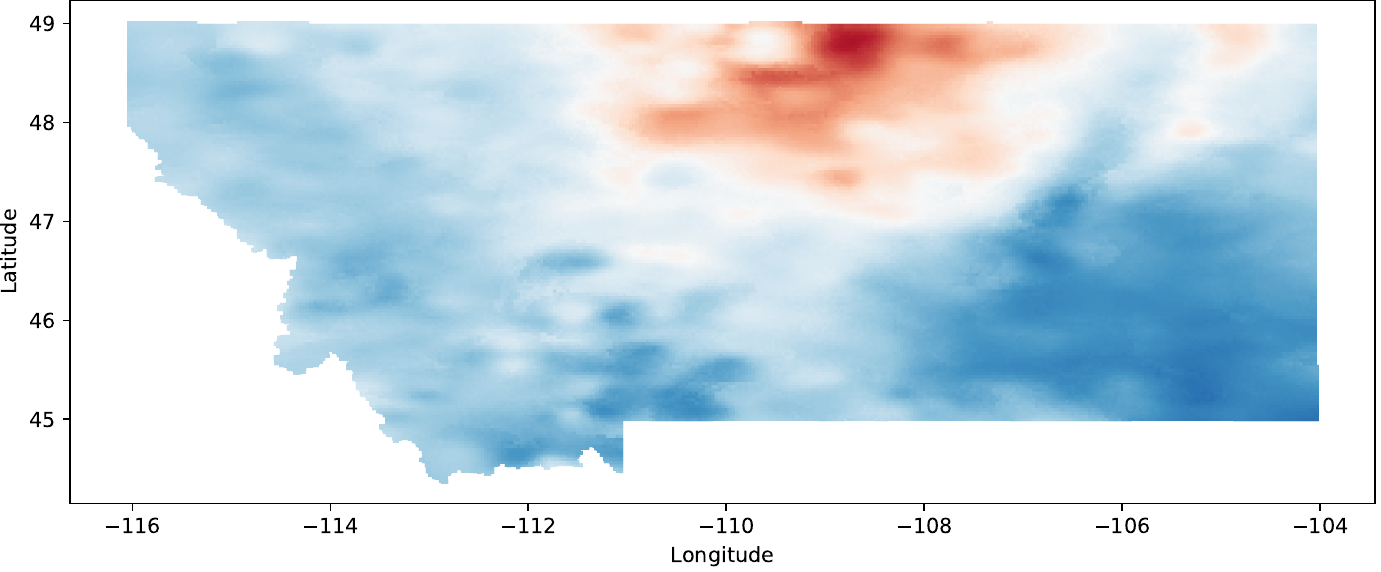}
        \caption{NNnGP}
        \label{fig:precipitation-NNnGP}
    \end{subfigure}\\[1ex]
    \includegraphics[width=0.4\textwidth]{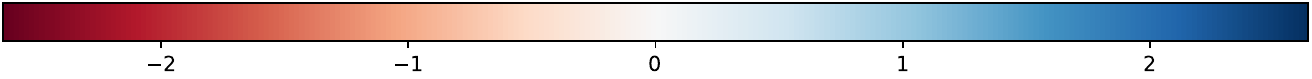}
    \caption{Precipitation anomaly and the predicted mean (for one replicate) in Montana, USA, during October, 2025.}
\end{figure}

The data came from the PRISM (Parameter-elevation Regressions on Independent Slopes Model) gridded monthly precipitation product, which utilizes a sophisticated climate model to interpolate real observations from a vast network of monitoring stations.\footnote{PRISM Group, Oregon State University, \url{https://prism.oregonstate.edu}, accessed 10 July 2026.}
Since precipitation is non-negative and typically right-skewed, we first applied a log-transformation to the total precipitation, then standardized each location to have a zero mean and unit variance based on historical records from the preceding $30$ years. 
The raw data, provided at an $800$-meter spatial resolution, was down-sampled to around $40,000$ spatial coordinates to facilitate computation while preserving the main spatial structure.\footnote{The down-sampling was implemented by regular spatial block aggregation, grouping nearby latitude-longitude grid cells into blocks and representing each block by the average location and standardized log precipitation anomaly value.}
The resulting precipitation anomaly data are shown in Figure~\ref{fig:precipitation-truth}.
The anomaly field exhibits pronounced spatial heterogeneity, characterized by severe, localized dry and wet extremes particularly concentrated in the northern and southeastern sectors of the state, respectively. 

We conducted $10$ replicate experiments. 
In each replicate, $800$ observations were randomly selected as the training set, while the remaining observations were held out for prediction.
We compared the proposed NNnGP with NNGP and NNMP. 
The NNnGP is trained using MAF-based VI similar to Section~\ref{sec:simulation}, with the number of hidden units per layer increased to $1024$.
Figures \ref{fig:precipitation-NNGP}-\ref{fig:precipitation-NNnGP} show the prediction means of the three methods for one replicate.
Visually, the prediction map of NNGP appears excessively smooth and blurry.
While NNMP captures slightly more local variation, it still over-smooths distinct spatial features.
In comparison, NNnGP produces a prediction map with sharper boundaries and finer textures that more closely resemble the ground truth.
Despite the visual improvements, the global prediction metrics (Table~\ref{tab:precipitation-global-metrics}), including RSR, CRPS, and overall credible interval coverage and width, are highly similar across all three methods.
One possible reason for the parity is that the global metrics are dominated by the majority of spatial locations exhibiting normal, non-extreme variations, where standard GP kriging suffices.

\begin{table}[tb]
\centering
\scriptsize
\caption{\small Global prediction performance for NNnGP, NNGP, and NNMP on the Montana precipitation anomaly data. Values are reported as mean (standard deviation) across 10 replicates.}
\label{tab:precipitation-global-metrics}
\begin{tabular}{lcccc}
\toprule
Method & RSR\% & CRPS & CI cover\% & CI width \\
\midrule
NNnGP & 23.21 (0.82) & 0.09 (0.00) & 95.03 (0.78) & 0.69 (0.03) \\
NNGP  & \textbf{21.43 (0.71)} & \textbf{0.08 (0.00)} & 95.04 (0.60) & \textbf{0.66 (0.02)} \\
NNMP  & 22.91 (0.58) & 0.09 (0.00) & \textbf{97.01 (0.49)} & 0.78 (0.02) \\
\bottomrule
\end{tabular}
\end{table}

Because precipitation anomalies are often most scientifically important in the tails, we separately evaluated tail prediction performance using multiple metrics, including the Brier Skill Score \parencite[BSS,][]{wilks2011statistical}, threshold-weighted CRPS \parencite[twCRPS,][]{gneiting2011comparing}, and extreme interval coverage and width.
The BSS evaluates the accuracy of predicted tail-event probabilities relative to empirical climatology,
with larger values indicating greater improvement over the climatological baseline.
Here, a tail event is defined by whether the anomaly exceeds a given lower- or upper-tail threshold.
The twCRPS assesses the accuracy of the predictive distribution within a specified tail region, with smaller values indicating better probabilistic forecasts in the tails.
Extreme interval coverage and width further evaluate whether the predictive intervals for extreme observations
are well calibrated and appropriately sharp.
The thresholds for tail events and extreme observations are set to the corresponding left and right quantiles of the standard normal distribution
over a range of tail probability levels.
Formal definitions of these metrics are provided in Supplement Section~\ref{sec:metrics-extreme}.

\begin{figure}[htbp]
    \centering
    %--------------------- SECOND ROW --------------
    \begin{subfigure}[b]{0.49\textwidth}
        \centering
        \includegraphics[width=\textwidth]{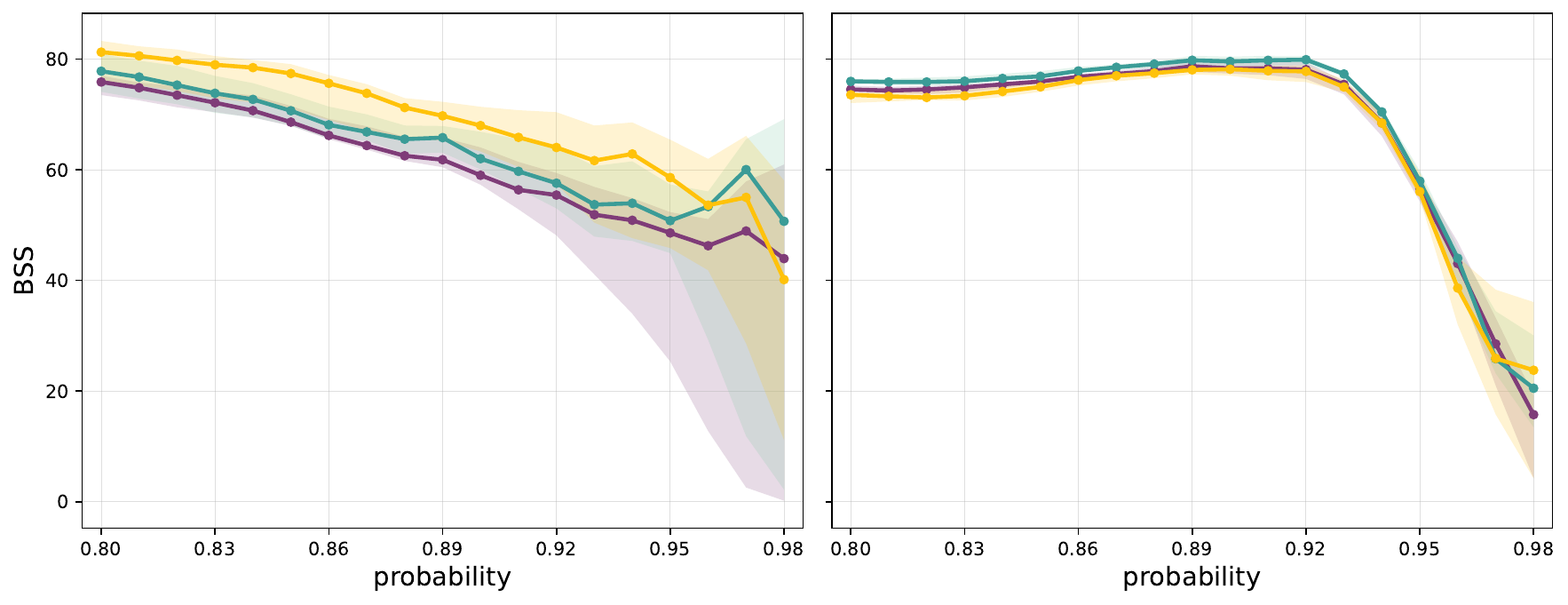}
        \caption{BSS (the higher the better)}
    \end{subfigure}\hfill
    \begin{subfigure}[b]{0.49\textwidth}
        \centering
        \includegraphics[width=\textwidth]{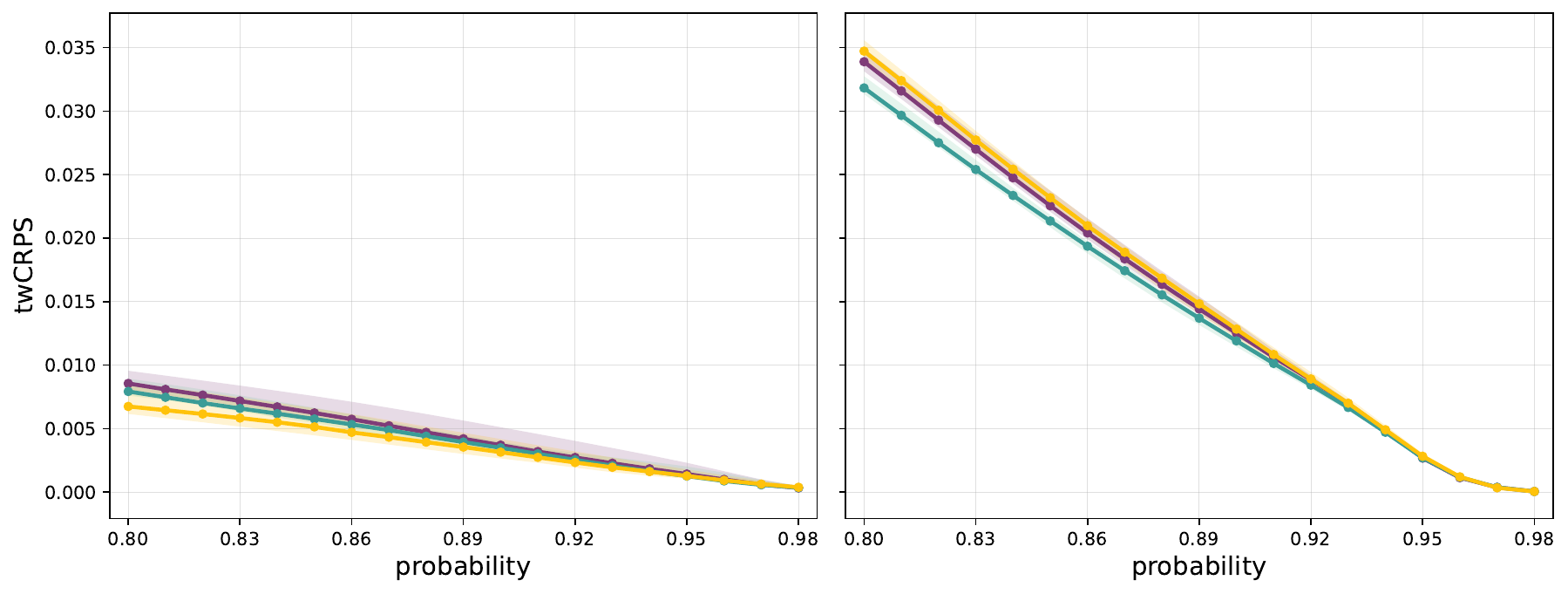}
        \caption{twCRPS (the lower the better)}
    \end{subfigure}\vspace{.05in}\hfill
    %--------------------- SECOND ROW --------------
    \begin{subfigure}[b]{0.49\textwidth}
        \centering
        \includegraphics[width=\textwidth]{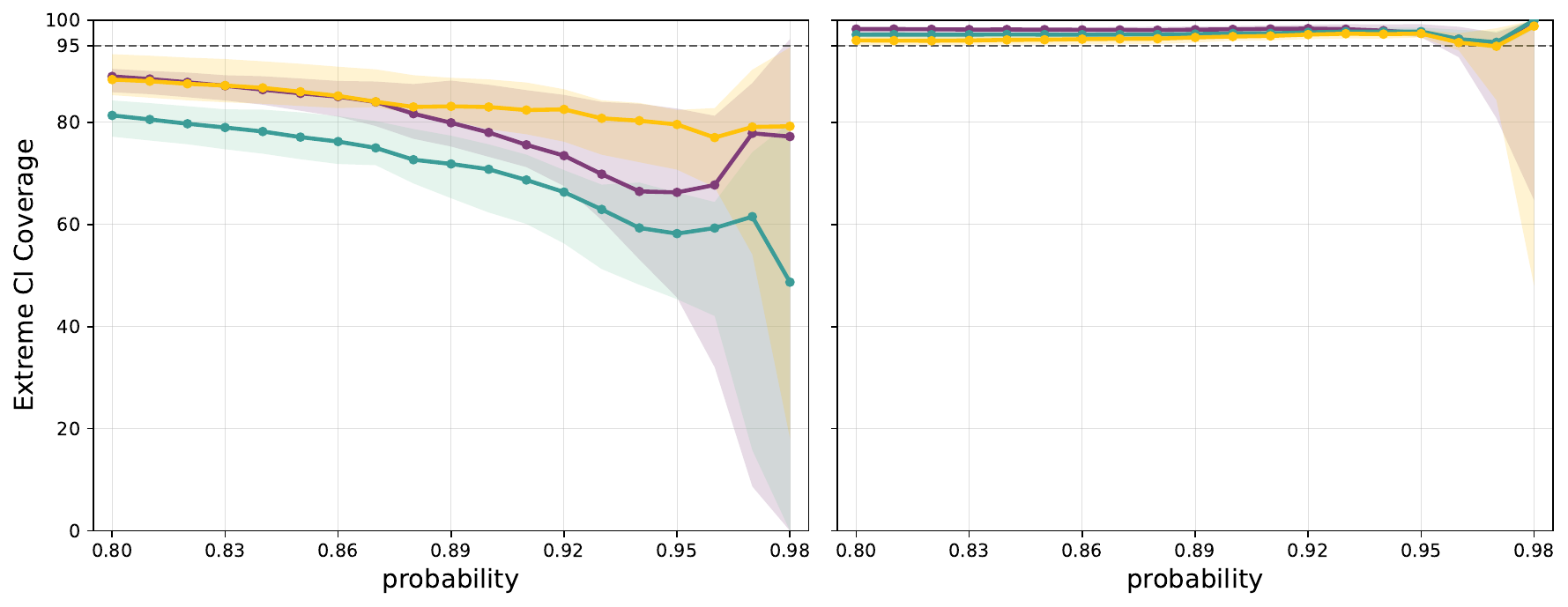}
        \caption{Extreme $95\%$ CI coverage}
    \end{subfigure}\hfill
    \begin{subfigure}[b]{0.49\textwidth}
        \centering
        \includegraphics[width=\textwidth]{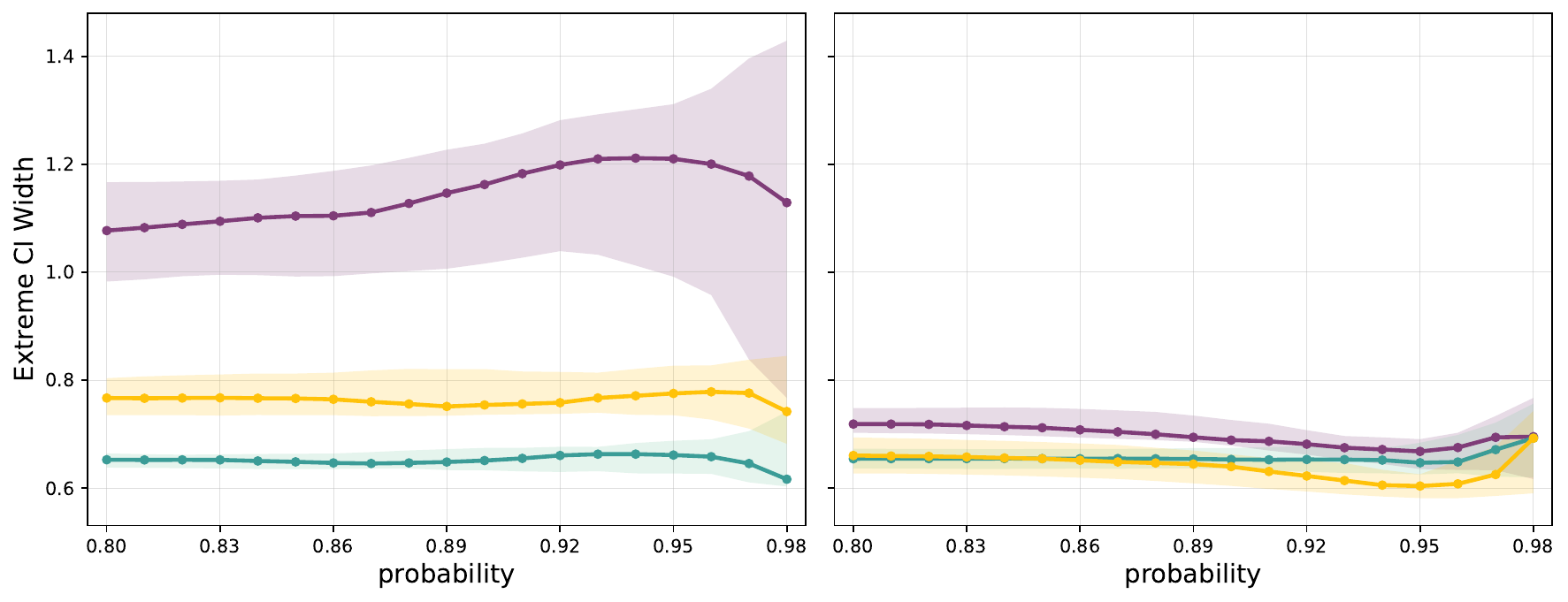}
        \caption{Extreme $95\%$ CI Width}
    \end{subfigure}\hfill\\[1ex]
    \includegraphics[width=0.3\textwidth]{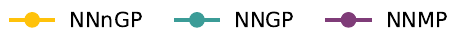}
    \caption{Tail prediction metrics of three models for a sequence of tail probability levels from $0.80$ to $0.98$. For each sub-figure, the left and right plots correspond to the left and right tails, respectively. Solid lines represent the median over $10$ replicates, and the shaded bands mark the $10$th and $90$th quantiles.}
    \label{fig:precipitation-tail-prediction}
\end{figure}

As shown in Figure~\ref{fig:precipitation-tail-prediction}, there is a significant asymmetry in prediction performance between the two tails.
In predicting the left tail (representing extreme drought conditions), NNnGP outperforms both NNGP and NNMP, delivering higher BSS, lower twCRPS, and better extreme coverage without producing excessively wide intervals.
In predicting the right tail (representing extremely heavy precipitation events), all three methods exhibit comparable and more desirable performance than in the left tail case.
This example demonstrates that as a flexible and robust modeling framework, NNnGP can extract complex, non-linear spatial dependencies, mitigating the underestimation issue of extreme events for traditional GP kriging.
Meanwhile, the introduction of the non-linear conditional mean does not induce overfitting, thereby preserving global predictive accuracy even in sub-regions where strong non-linearity is absent.
However, while NNnGP outperforms the baselines, the predictive performance for all methods declines when moving further into the extreme tails (i.e., with smaller tail probabilities). 
We will further discuss the methodological implications of this observation in Section~\ref{sec:conclusion}.

\section{Conclusion}\label{sec:conclusion}

In this article, we have proposed the nearest-neighbor non-Gaussian process (NNnGP), a flexible framework for geostatistical modeling. 
By generalizing the conditional mean and variance functions within the Vecchia approximation, NNnGP extends the popular NNGP to capture complex, non-linear local dependencies. 
To ensure coherent spatial prediction from a single observation, we introduced a regularized homogeneity condition. 
We then developed a Bayesian non-parametric formulation for the conditional mean with a Gaussian process, which balances model interpretability and flexibility by directly enforcing desired properties such as the screening effect.
Closed-form probability densities are derived after marginalizing out the GP, which, combined with low-rank approximations, facilitates a normalizing flow-based variational inference for the NNnGP model.
As demonstrated in our numerical studies, NNnGP is a robust extension of NNGP, which achieves better global predictive accuracy than NNGP when there is strong non-linear local dependence, and stays on par with NNGP without overfitting when the non-linearity is absent.
The PRISM precipitation data further indicates that the non-linear conditional mean of NNnGP can mitigate the underestimation of extreme spatial events by traditional GP.
Despite these improvements, several methodological enhancements can be explored to increase both model flexibility and computational efficiency. 

First, while NNnGP improves tail predictions, the performance still declines as the events become exceptionally rare. 
This limitation inherently stems from the Gaussian conditionals.
One important extension of the current NNnGP framework is to use non-Gaussian distributions in (\ref{eq:conditional-unified}):
\begin{equation}\label{eq:conditional-unified-general}
    p(w(\bv)|\bw_{N(\bv)}) = p_0(w(\bv)|\mathbf{f}(\bw_{N(\bv)}, N(\bv),\bv)),
\end{equation}
for $\bv\in\cD\backslash\cS_{m+1}$.
Here, $p_0(w|\btheta)$ denotes a generic distribution with parameters $\btheta\in\Theta$, and $\mathbf{f}\colon \bbR^m\times \cD^{m}\times \cD\goto \Theta$ is a continuous mapping from the spatially varying arguments $(\bw_{N(\bv)}, N(\bv),\bv)$ to the parameter space.
The base distribution $p_0$ can encode prior information about the process, such as Student's $t$-distributions or generalized extreme value (GEV) distributions for modeling heavy tails and extremes, the skew-Gaussian distributions \parencite{azzalini2013skew} for modeling skewness, and the Beta distributions for modeling bounded measurements.
To avoid a case-by-case formulation of the mapping $\mathbf{f}$, highly expressive neural architectures, such as Graph Neural Networks (GNNs), could serve as universal approximations for these spatially varying parameters.

Second, NNnGP relies on a global regularized homogeneity assumption. 
This condition could be relaxed to better accommodate highly non-stationary spatial fields by partitioning the spatial domain into disjoint sub-regions, assuming localized homogeneity and fitting an independent NNnGP model within each sub-region. 
The partitioning idea has been widely applied to modeling non-stationary GPs; see \textcite{das2010block}, \textcite{lenzi2020improving}, and \textcite{luo2024nonstationary} for related work.

Third, the specific Bayesian non-parametric formulation studied in Section~\ref{sec:NNnGP-GP} faces scalability bottlenecks under massive datasets,
primarily due to the kernel matrix operations slowing down automatic differentiation and the MAF framework demanding $O(k^2)$ variational parameters.
Two separate approaches might be considered.
One approach, from the inference perspective, is to explicitly embed a conditional independence structure into the autoregressive flow, which exactly matches the DAG in the NNnGP prior on reference locations. 
As a result, the variational parameters can be drastically sparsified.
See \textcite{wehenkel2021graphical} for related work.
As an alternative to VI, neural simulation-based inference \parencite{zammit2025neural} may also be considered.
The other approach, from the modeling perspective, is to adopt highly expressive neural network architectures like GNNs in place of GPs, to model the conditional mean function or the spatially varying parameters $\mathbf{f}$ in (\ref{eq:conditional-unified-general}).
A neural-net formulation entirely bypasses the computational bottlenecks associated with kernel matrices, significantly increasing the model flexibility, while maintaining the interpretability of the NNnGP's probabilistic modeling.

Finally, it is also interesting to expand the NNnGP framework to accommodate broader classes of spatial data.
One natural extension is adapting the model for multivariate spatial responses, enabling the joint modeling of interacting non-Gaussian processes. 
Additionally, the current nearest neighbors are determined by the Euclidean distance, which is unsuitable for real-world applications with $3$-dimensional domains, like those involving altitude or ocean depth. 
In those problems, vertical variations are often governed by fundamentally different dynamics from planar variations.
Formulating an anisotropic NNnGP variant by adopting weighted distances---akin to Automatic Relevance Determination (ARD) kernels---might bridge this gap.

\printbibliography[title={References}]

\newpage
\appendix
\counterwithin{equation}{section}
\counterwithin{figure}{section}

\begin{center}
    \vspace*{2cm}
    {\LARGE Supplementary Materials for ``Nearest-Neighbor Non-Gaussian Processes for Geostatistical Modeling''\par}
    \vspace{.2in}
    
    Penghui Fu, Yuhan Dong, Jianhua Z. Huang\footnotemark[1]
\end{center}
\footnotetext[1]{Corresponding author: \href{mailto:jhuang@cuhk.edu.cn}{jhuang@cuhk.edu.cn}}

\begin{refsection}% to separate the references of main paper and the supplements
\section{Proofs}\label{sec:proofs}
\subsection{Proof of Proposition~\ref{pro:density-ws}}
Let $\bg_{\cS'} = (g_{m+1},\ldots,g_k)^\top = (g(\tilde\bw_i))^\top_{m+1\leq i \leq k}$ be the realizations of $g$ on inputs $\tilde\bw_{\cS'}$.
By the chain rule, we have
\begin{equation}\label{eq:density-ws-gs}
    \begin{split}
        p(\bw_\cS, \bg_{\cS'}) &= p(\bw_{1:m})\cdot p(\bw_{\cS'}, \bg_{\cS'}|\bw_{1:m}) \\
        &= p(\bw_{1:m})\cdot \left[\prod_{i=m+1}^k p(w_i, g_i|\bw_{1:(i-1)}, \bg_{(m+1):(i-1)})\right] \\
        &= p(\bw_{1:m})\cdot \left[\prod_{i=m+1}^k p(g_i|\bw_{1:(i-1)}, \bg_{(m+1):(i-1)})p(w_i|\bw_{1:(i-1)},\bg_{(m+1):i})  \right] \\
        &= p(\bw_{1:m})\cdot \prod_{i=m+1}^k \left [p(g_i|\bg_{(m+1):(i-1)};\tilde\bw_i,\tilde\bw_{(m+1):(i-1)})N(w_i|h_i + \tau_i g_i, F_i) \right] \\
        &= N(\bw_{1:m}|\bzero, \bC_{mm})
        N(\bg_{\cS'}|\bzero, \bR_{\tilde\bw_{\cS'}}) N(\bw_{\cS'}|\bh_{\cS'} + \bD_{\btau_{\cS'}}\bg_{\cS'}, \bD_{\bF_{\cS'}}),
    \end{split}
\end{equation}
where in the penultimate equation, we use the semi-colon notation to clarify the input locations of the corresponding function values, such that $p(g_i|\bg_{(m+1):(i-1)};\tilde\bw_i,\tilde\bw_{(m+1):(i-1)})$ indicates that the input locations for $g_i$ and $\bg_{(m+1):(i-1)}$ are $\tilde\bw_i$ and $\tilde\bw_{(m+1):(i-1)})$, respectively;
in the last equation, we have applied the chain rule in reverse to $N(\bg_{\cS'}|\bzero, \bR_{\tilde\bw_{\cS'}})$:
\begin{equation*}
    N(\bg_{\cS'}|\bzero, \bR_{\tilde\bw_{\cS'}}) = \prod_{i=m+1}^k p(g_i|\bg_{(m+1):(i-1)};\tilde\bw_i,\tilde\bw_{(m+1):(i-1)}).
\end{equation*}
After rearrangement, (\ref{eq:density-ws-gs}) can be put into
\begin{equation}\label{eq:eq:density-ws-gs-2}
     p(\bw_\cS, \bg_{\cS'}) = p_{\mathrm{NNGP}}(\bw_\cS) \frac{1}{|\bR_{\tilde\bw_{\cS'}}|}\mathrm{exp}\left\{\frac{1}{2}\ba^\top \bG\ba\right\}\mathrm{exp}\left\{-\frac{1}{2}(\bg-\bG\ba)^\top\bG^{-1}(\bg-\bG\ba)\right\},
\end{equation}
where
\begin{equation*}
    \bG^{-1} = \bD_{\btau_{\cS'}}\bD_{\bF_{\cS'}}^{-1}\bD_{\btau_{\cS'}} + \bR_{\tilde\bw_{\cS'}}^{-1}, \quad \ba = \bD_{\btau_{\cS'}}\bD_{\bF_{\cS'}}^{-1}(\bw_{\cS'} - \bh_{\cS'}),
\end{equation*}
and 
\begin{equation*}
    \begin{split}
        p_{\mathrm{NNGP}}(\bw_\cS) &= N(\bw_{1:m}|\bzero, \bC_{mm})\frac{1}{\sqrt{|\bD_{\bF_{\cS'}}|}}\mathrm{exp}\left\{-\frac{1}{2}\left((\bw_{\cS'} - \bh_{\cS'})^\top\bD_{\bF_{\cS'}}^{-1}(\bw_{\cS'} - \bh_{\cS'})\right)\right\} \\
        &= N(\bw_{1:m}|\bzero, \bC_{mm})\prod_{i=m+1}^k N(w_i|h_i,F_i) = \prod_{i=1}^k N(w_i|h_i,F_i).
    \end{split}
\end{equation*}
Collecting the terms related to $\bg_{\cS'}$ in (\ref{eq:eq:density-ws-gs-2}) yields 
\begin{equation}\label{eq:density-gs-given-ws}
    p(\bg_{\cS'}|\bw_\cS) = N(\bg_{\cS'}|\bG\ba, \bG).
\end{equation}
Then, we have
\begin{equation*}
    \begin{split}
        p(\bw_\cS) = p(\bw_\cS,\bg_{\cS'})/p(\bg_{\cS'}|\bw_\cS) = p_{\mathrm{NNGP}}(\bw_\cS)\sqrt{\frac{|\bG|}{|\bR_{\tilde\bw_{\cS'}}|}}\mathrm{exp}\left\{\frac{1}{2} \ba^\top\bG\ba\right\},
    \end{split}
\end{equation*}
which concludes the proof.

\subsection{Proof of Proposition~\ref{pro:density-wu-given-ws}}
Let $\bg_\cU = (g(\tilde\bw_{\bu_1}),\ldots,g(\tilde\bw_{\bu_r}))^\top$.
Given $\bg_\cU$ and $\bw_\cS$, $w(\bu_1),\ldots,w(\bu_r)$ are independent Gaussian variables with mean $h_{\bu_i} + \tau_{\bu_i} g(\tilde\bw_{\bu_i})$ and variance $F_{\bu_i}$, for $i=1,\ldots,r$.
Therefore, 
\begin{equation*}
    \begin{split}
        p(\bw_\cU,\bg_\cU|\bw_\cS,\bg_{\cS'}) &= p(\bg_\cU|\bw_\cS,\bg_{\cS'})p(\bw_\cU|\bw_\cS,\bg_\cU) \\
        &= p(\bg_\cU|\bg_{\cS'};\tilde\bw_\cU,\tilde\bw_{\cS'})N(\bw_\cU|\bh_\cU + \bD_{\btau_\cU}\bg_\cU, \bD_{\bF_\cU}) \\
        &= N(\bg_\cU|\widetilde\bB_{\tilde\bw_\cU,\tilde\bw_{\cS'}}\bg_{\cS'}, \bR_{\tilde\bw_\cU|\tilde\bw_{\cS'}})N(\bw_\cU|\bh_\cU + \bD_{\btau_\cU}\bg_\cU, \bD_{\bF_\cU}).
    \end{split}
\end{equation*}
Marginalizing out $\bg_\cU$ in the above equation yields
\begin{equation}\label{eq:density-wu-given-ref}
    p(\bw_\cU|\bw_\cS,\bg_{\cS'}) = 
    N(\bw_\cU|\bh_\cU + \bD_{\btau_\cU}\widetilde\bB_{\tilde\bw_\cU,\tilde\bw_{\cS'}}\bg_{\cS'},\,\,\bD_{\bF_\cU} + \bD_{\btau_\cU}\bR_{\tilde\bw_\cU|\tilde\bw_{\cS'}}\bD_{\btau_\cU}).
\end{equation}
By (\ref{eq:density-gs-given-ws}) and (\ref{eq:density-wu-given-ref}), we have
\begin{equation*}
    \begin{split}
        p(\bw_\cU|\bw_\cS) &= \int p(\bg_{\cS'},\bw_\cU|\bw_\cS)\,\dif\bg_{\cS'} \\
        &= \int p(\bg_{\cS'}|\bw_\cS)p(\bw_\cU|\bw_\cS,\bg_{\cS'})\,\dif\bg_{\cS'} \\
        &= \int N(\bg_{\cS'}|\bG\ba, \bG)N(\bw_\cU|\bh_\cU + \bD_{\btau_\cU}\widetilde\bB_{\tilde\bw_\cU,\tilde\bw_{\cS'}}\bg_{\cS'},\,\,\bD_{\bF_\cU} + \bD_{\btau_\cU}\bR_{\tilde\bw_\cU|\tilde\bw_{\cS'}}\bD_{\btau_\cU})\,\dif\bg_{\cS'} \\
        &= N(\bw_\cU|\bh_\cU + \bD_{\btau_\cU}\widetilde\bB_{\tilde\bw_\cU,\tilde\bw_{\cS'}}\bG\ba,\,\,\bD_{\bF_\cU} + \bD_{\btau_\cU}(\bR_{\tilde\bw_\cU|\tilde\bw_{\cS'}} + \widetilde\bB_{\tilde\bw_\cU,\tilde\bw_{\cS'}}\bG\widetilde\bB_{\tilde\bw_{\cS'},\tilde\bw_\cU})\bD_{\btau_\cU}),
    \end{split}
\end{equation*}
which concludes the proof.

% For convenience, we take $g_1 = g_2 = \cdots = g_m = 0$, and assume that $p(w_i|w_{1:i-1}) = N(w_i|h_i + \tau_i g_i, F_i) = N(w_i|h_i, F_i)$ holds for $i=1,\ldots,m$. 
% See Figure \ref{fig:graph-rep} for a graphical representation of $\bw_\cS$.
% \begin{figure}[t]
%   \centering
%   \input{nongau-graph-rep}   
%   \caption{The graphical representation of $\bw_\cS$ for TNP-GP.}
%   \label{fig:graph-rep}
% \end{figure}

\section{Low-rank Approximations}\label{sec:low-rank}
\subsection{Overview}
In this section, we calculate (\ref{eq:density-ws}) and (\ref{eq:density-wu-given-ws}) after replacing the original kernel $R$ by $\tilde R$ in (\ref{eq:FIC}).
To simplify notations, we use $\bW$ and $\bV$ as shorthands for $\tilde\bw_{\cS'}$ and $\tilde\bw_\cU$, respectively. 
% See Table~\ref{tab:complexity} for a summary of the computational and storage costs of major quantities.\footnote{The computational cost of a quantity includes the costs of calculating all required intermediate quantities. For diagonal matrices, it suffices to store their diagonal entries.}
For inference of $\bw_{\cS}$, the overall computational and storage costs of evaluating (\ref{eq:density-ws}) are $O(k(\tilde m^2+m^3))$ and $O(k\tilde m)$, respectively.
For prediction, the overall computational and storage costs of evaluating the mean and covariance matrix in (\ref{eq:density-wu-given-ws}) are $O(k(\tilde m r + \tilde m^2 + r^2 + m^3))$ and $O(r^2)$, respectively. 
Sampling from the joint normal further takes $O(r^3)$, which is intractable for large $r$.
However, if location-wise rather than joint predictions are of interest, it suffices to compute the diagonal elements of the covariance matrix (i.e., the location-wise variances), which costs $O(kr\tilde m)$, and sample $r$ locations independently.
% An alternative approach is to partition the $r$ locations into mini-batches and sample each batch independently. 
% Then the covariance structure within each batch can be preserved.

\subsection{Computation Details}
For (\ref{eq:density-ws}), by (\ref{eq:FIC}), we have 
$$
\widetilde\bR_\bW = \bQ_\bW + \diag(\bR_\bW - \bQ_\bW),
$$
where $\bQ_\bW = \bR_{\bW\bZ}\bR_{\bZ}^{-1}\bR_{\bZ\bW}$.
Let $\bD_1 = \diag(\bR_\bW - \bQ_\bW) = \bI - \diag(\bQ_\bW)$. By the matrix determinant lemma, 
$$
|\widetilde\bR_\bW| = |\bD_1||\bR_\bZ|^{-1}|\bR_\bZ + \bR_{\bZ\bW}\bD_1^{-1}\bR_{\bW\bZ}|.
$$
By Sherman--Morrison--Woodbury,
$$
\widetilde\bR_\bW^{-1} = \bD_1^{-1} - \bD_1^{-1}\bR_{\bW\bZ}(\bR_{\bZ} + \bR_{\bZ\bW}\bD_1^{-1}\bR_{\bW\bZ})^{-1}\bR_{\bZ\bW}\bD_1^{-1}.
$$
Let $\bC_1 = \bR_{\bZ} + \bR_{\bZ\bW}\bD_1^{-1}\bR_{\bW\bZ}$ and $\bD_2^{-1}=\bD_{\btau_{\cS'}}\bD_{\bF_{\cS'}}^{-1}\bD_{\btau_{\cS'}}$. We have 
$$
\bG^{-1} = \bD_2^{-1} + \widetilde\bR_\bW^{-1} = \bD_1^{-1} + \bD_2^{-1} - \bD_1^{-1}\bR_{\bW\bZ}\bC_1^{-1}\bR_{\bZ\bW}\bD_1^{-1}.
$$
Let $\bD_3^{-1} = \bD_1^{-1} + \bD_2^{-1}$ and $\bC_2 = \bC_1 - \bR_{\bZ\bW}\bD_1^{-1}\bD_3\bD_1^{-1}\bR_{\bW\bZ}$.
Then
$$
%|\bG|^{-1} = |\bD_3|^{-1}|\bC_1|^{-1}|\bC_1 - \bR_{\bZ\bW}\bD_1^{-1}\bD_3\bD_1^{-1}\bR_{\bW\bZ}|,
|\bG|^{-1} = |\bD_3|^{-1}|\bC_1|^{-1}|\bC_2|,
$$
and 
$$
\bG = \bD_3 + (\bD_3\bD_1^{-1}\bR_{\bW\bZ})\bC_2^{-1}(\bR_{\bZ\bW}\bD_1^{-1}\bD_3).
$$
Utilizing the low-rank structure of $\bG$, $\bG\ba$ can be efficiently evaluated as
$$
\bG\ba = \bD_3\ba + (\bD_3\bD_1^{-1}\bR_{\bW\bZ})\bC_2^{-1}(\bR_{\bZ\bW}\bD_1^{-1}\bD_3\ba),
$$
by first calculating matrix products in the parentheses. 

For (\ref{eq:density-wu-given-ws}), the main bottleneck in the mean term is $\widetilde\bB_{\bV,\bW}\bG\ba$. 
We first notice that $\widetilde\bR_{\bV\bW} = \bQ_{\bV\bW} = \bR_{\bV\bZ}\bR_\bZ^{-1}\bR_{\bZ\bW}$ can be evaluated in $O(k(\tilde m r + \tilde m^2))$. 
Then,
\begin{equation*}
    \begin{split}
        \widetilde\bB_{\bV,\bW} &= \widetilde\bR_{\bV\bW}\widetilde\bR_\bW^{-1} = \widetilde\bR_{\bV\bW}(\bD_1^{-1} - \bD_1^{-1}\bR_{\bW\bZ}\bC_1^{-1}\bR_{\bZ\bW}\bD_1^{-1})\\
        & = \widetilde\bR_{\bV\bW}\bD_1^{-1} - (\widetilde\bR_{\bV\bW}\bD_1^{-1}\bR_{\bW\bZ})\bC_1^{-1}\bR_{\bZ\bW}\bD_1^{-1},
    \end{split}    
\end{equation*}
which can be evaluated in $O(k(\tilde m r + \tilde m^2))$ by first computing the product in the parenthesis.
After that, the quantity $\widetilde\bB_{\bV,\bW}(\bG\ba)$ can be efficiently evaluated, since $\bG\ba$ has already been calculated. 

For the covariance matrix, the main bottleneck is in evaluating $\widetilde\bR_{\bV|\bW} + \widetilde\bB_{\bV,\bW}\bG\widetilde\bB_{\bW,\bV}$.
Let $\bD_4 = \diag(\bR_\bV - \bQ_\bV) = \bI - \diag(\bQ_\bV)$.
We have
\begin{equation*}
    \begin{split}
        \widetilde\bR_{\bV|\bW} &= \widetilde\bR_{\bV\bV} - \widetilde\bR_{\bV\bW}\widetilde\bR_\bW^{-1}\widetilde\bR_{\bW\bV}\\
        &= (\bR_{\bV\bZ}\bR_{\bZ}^{-1}\bR_{\bZ\bV} + \bD_4) - \widetilde\bR_{\bV\bW}(\bD_1^{-1} - \bD_1^{-1}\bR_{\bW\bZ}\bC_1^{-1}\bR_{\bZ\bW}\bD_1^{-1})\widetilde\bR_{\bW\bV} \\
        &= (\bR_{\bV\bZ}\bR_{\bZ}^{-1}\bR_{\bZ\bV} + \bD_4) - (\widetilde\bR_{\bV\bW}\bD_1^{-1}\widetilde\bR_{\bW\bV}) + (\widetilde\bR_{\bV\bW}\bD_1^{-1}\bR_{\bW\bZ})\bC_1^{-1}(\bR_{\bZ\bW}\bD_1^{-1}\widetilde\bR_{\bW\bV}).
    \end{split}
\end{equation*}
Similarly,
\begin{equation*}
    \begin{split}
        \widetilde\bB_{\bV,\bW}\bG\widetilde\bB_{\bW,\bV} &= \widetilde\bB_{\bV,\bW}[\bD_3 + (\bD_3\bD_1^{-1}\bR_{\bW\bZ})\bC_2^{-1}(\bR_{\bZ\bW}\bD_1^{-1}\bD_3)]\widetilde\bB_{\bW,\bV}  \\
        &= \widetilde\bB_{\bV,\bW}\bD_3\widetilde\bB_{\bW,\bV} + (\widetilde\bB_{\bV,\bW}\bD_3\bD_1^{-1}\bR_{\bW\bZ})\bC_2^{-1}(\bR_{\bZ\bW}\bD_1^{-1}\bD_3\widetilde\bB_{\bW,\bV}).
    \end{split}
\end{equation*}
Explicitly calculating $\widetilde\bR_{\bV|\bW} + \widetilde\bB_{\bV,\bW}\bG\widetilde\bB_{\bW,\bV}$ is costly for large $r$. 
Nonetheless, their diagonal elements can be efficiently calculated in $O(kr\tilde m)$ by noticing that
\begin{equation*}
    \begin{split}
        \diag(\widetilde\bR_{\bV|\bW}) &= (\bR_{\bV\bZ}\bR_{\bZ}^{-1/2})^{\odot 2}\mathbf{1} + \bD_4 - (\widetilde\bR_{\bV\bW}\bD_1^{-1/2})^{\odot 2}\mathbf{1} + (\widetilde\bR_{\bV\bW}\bD_1^{-1}\bR_{\bW\bZ}\bC_1^{-1/2})^{\odot 2}\mathbf{1},
    \end{split}
\end{equation*}
and 
\begin{equation*}
    \begin{split}
        \diag(\widetilde\bB_{\bV,\bW}\bG\widetilde\bB_{\bW,\bV}) &= (\widetilde\bB_{\bV,\bW}\bD_3^{1/2})^{\odot 2}\mathbf{1} + (\widetilde\bB_{\bV,\bW}\bD_3\bD_1^{-1}\bR_{\bW\bZ}\bC_2^{-1/2})^{\odot 2}\mathbf{1},
    \end{split}
\end{equation*}
where we have used the fact that for any matrices $\bA$, the diagonal elements of $\bA\bA^\top$ can be calculated as $\bA^{\odot 2}\mathbf{1}:=(\bA\odot\bA)\mathbf{1}$.

\section{Inference when \texorpdfstring{$\cT\nsubseteq\cS$}{T not in S}}\label{sec:prediction-T-notsubsetof-S}
We discuss the inference when the reference points are chosen such that $\cT\nsubseteq\cS$.
Denote $\cU=\cT\backslash\cS$ as the non-reference locations where we have observations. 
The Bayesian hierarchical model is
\begin{equation}\label{eq:hierarchical-model-nonsubset}
    \begin{split}
        p(\by_\cT|\bw_{\cS\cup\cT}) &= N(\by_\cT|\bX_\cT\bbeta + \bw_\cT, \sigma^2_\epsilon\bI), \\
        p(\bw_{\cS\cup\cT}) &= p(\bw_{\cS\cup\cU}) = p(\bw_\cS)p(\bw_\cU|\bw_\cS),
    \end{split}
\end{equation}
where $p(\bw_\cS)$ is given by (\ref{eq:density-ws}) and $p(\bw_\cU|\bw_\cS)$ is given by (\ref{eq:density-wu-given-ws}).
Compared with (\ref{eq:hierarchical-model-subset}), (\ref{eq:hierarchical-model-nonsubset}) involves additional latent variables $\bw_\cU$.
Although $p(\bw_\cS)$ can be evaluated efficiently, $p(\bw_\cU|\bw_\cS)$ costs $O(k(\tilde m r + \tilde m^2 + r^2 + m^3)+r^3)$ with $r = |\cU|$, which is intractable when $r$ is large. 
Likelihood-free methods may be used to handle this scenario.
After the inference, predictions at unobserved locations and their uncertainty quantification can be computed similarly to the case $\cT\subseteq\cS$. 

Let $\by_{\tilde\cV}$ denote the responses at locations $\tilde\cV$. 
We have the following procedure to sample $\by_{\tilde\cV}$.

\begin{minipage}{0.9\linewidth}
\begin{algorithm}[H]
\caption{Making predictions (when $\cT\nsubseteq\cS$)}
%\footnotesize
\begin{algorithmic}
\STATE {\textbf{Input}} Model parameters $(\btheta,\bbeta, \sigma^2_\epsilon)$.
\REPEAT
\STATE Sample $\bw_{\cS\cup\cT}\sim p(\bw_{\cS\cup\cT}|\by_\cT)$.
\STATE For locations $\cV_0 = \tilde\cV\cap (\cS\cup\cT)$, extract $\bw_{\cV_0}$.
\STATE For locations $\cV = \tilde\cV\backslash (\cS\cup\cT)$, sample $\bw_\cV\sim p(\bw_\cV|\bw_{\cS\cup\cT})$.
\STATE Combine $\bw_{\tilde\cV} = (\bw_{\cV_0}, \bw_\cV)$.
\STATE Sample $\by_{\tilde\cV}\sim N(\bX_{\tilde\cV}\bbeta + \bw_{\tilde\cV}, \sigma^2_\epsilon\bI)$.
\UNTIL{A desired number of samples are obtained.}
\end{algorithmic}
\end{algorithm}
\end{minipage}
\vspace{.2in}\\
In the above procedure, for locations $\cV = \tilde\cV\backslash (\cS\cup\cT)$, $p(\bw_\cV|\bw_{\cS\cup\cT})$ is jointly Gaussian.
To be more specific, let $\cU = \cT\backslash\cS$ and $\cZ=\cV\cup\cU$. Then, $p(\bw_\cV|\bw_{\cS\cup\cT}) = p(\bw_\cV|\bw_{\cS\cup\cU})$.
By Proposition~\ref{pro:density-wu-given-ws}, we have
$p(\bw_\cZ|\bw_\cS) = N(\bw_\cZ|\bmu_\cZ, \bXi_{\cZ,\cZ})$,
where $\bmu_\cZ$ and $\bXi_{\cZ,\cZ}$ are the mean vector and covariance matrix corresponding to the mean function (\ref{eq:wu-given-ws-mean}) and covariance function (\ref{eq:wu-given-ws-covariance}) of the conditioning process $w(\cdot)|\bw_\cS$, respectively.
Therefore, $p(\bw_\cV|\bw_{\cS\cup\cU})  = N(\bw_\cV|\bmu_\cV + \bXi_{\cV\cU}\bXi_{\cU\cU}^{-1}(\bw_\cU-\bmu_\cU), \bXi_{\cV\cV} - \bXi_{\cV\cU}\bXi_{\cU\cU}^{-1}\bXi_{\cU\cV})$.

\section{Flow-based Variational Inference}\label{sec:flow-vi}

In this section, we provide a brief overview of the flow-based VI used in our spatial regression model.
Our goal is to estimate the unknown model hyperparameters $\bTheta = (\btheta, \bbeta, \sigma^2_\epsilon)$, and approximate the intractable posterior of the latent spatial variables, $p(\bw_\cS|\by_\cT, \bTheta)$, using a tractable variational density $q_\phi(\bw_\cS)$ parameterized by $\phi$.

To construct a highly flexible non-Gaussian distribution, $q_\phi(\bw_\cS)$ is modeled via a normalizing flow. 
Specifically, $\bw_\cS$ is generated by applying a sequence of differentiable and invertible transformations $T_\phi = T_K \circ \cdots \circ T_1$ to a base noise variable $\bepsilon \sim p_\text{base}(\bepsilon)$, where $p_\text{base}$ is typically a standard multivariate normal distribution. 
By the change-of-variables formula, the variational density is
\begin{equation*}
    q_\phi(\bw_\cS) = p_\text{base}(\bepsilon) \left|J_{T_\phi}(\bepsilon)\right|_{+}^{-1} \Big|_{\bepsilon = T_\phi^{-1}(\bw_\cS)},
\end{equation*}
where $J_{T_\phi}(\bepsilon)$ is the Jacobian matrix, and $\left|J_{T_\phi}(\bepsilon)\right|_{+}$ denotes the absolute value of its determinant.
The model parameters $\bTheta$ and variational parameters $\phi$ are jointly estimated by maximizing the Evidence Lower Bound (ELBO):
\begin{equation}\label{eq:ELBO}
    \begin{split}
        L(\bTheta, \phi) &= \mathbb{E}_{\bw_\cS \sim q_\phi(\bw_\cS)} \left[\log p(\by_\cT, \bw_\cS\mid \bTheta) - \log q_\phi(\bw_\cS) \right] \\
        &= \mathbb{E}_{\bepsilon \sim p_\text{base}(\bepsilon)}\left[\log p\big(\by_\cT, \bw_\cS = T_\phi(\bepsilon) \mid \bTheta\big) - \log p_\text{base}(\bepsilon) + \log \left|J_{T_\phi}(\bepsilon)\right|_{+}\right],
    \end{split}
\end{equation}
where in the last equation we made a change of variables, known as the reparameterization trick.

For gradient-based optimization, the gradients of (\ref{eq:ELBO}) with respect to the hyperparameters $\bTheta$ and flow parameters $\phi$ can be estimated via Monte Carlo sampling:
\begin{equation*}
    \begin{split}
        \nabla_{\bTheta} L(\bTheta, \phi) = \bbE_{\bepsilon \sim p_\text{base}(\bepsilon)} [\nabla_{\bTheta} \log p(\by_\cT, \bw_\cS = T_\phi(\bepsilon) \mid \bTheta) ] \approx \frac{1}{N}\sum_{j=1}^N \nabla_{\bTheta} \log p\big(\by_\cT, \bw_\cS^{(j)} \mid \bTheta\big),
    \end{split}
\end{equation*}
and
\begin{equation*}
    \begin{split}
        \nabla_\phi L(\bTheta, \phi) &= \bbE_{\bepsilon \sim p_\text{base}(\bepsilon)}\left[\nabla_\phi  T_\phi(\bepsilon)\nabla_{\bw_\cS}\log p\big(\by_\cT, \bw_\cS \mid \bTheta\big)\Big|_{\bw_\cS=T_\phi(\bepsilon)} + \nabla_\phi\log \left|J_{T_\phi}(\bepsilon)\right|_+ \right]\\
        &\approx \frac{1}{N}\sum_{j=1}^N \left[ \nabla_\phi T_\phi(\bepsilon^{(j)}) \nabla_{\bw_\cS}\log p\big(\by_\cT, \bw_\cS \mid \bTheta\big)\Big|_{\bw_\cS=T_\phi(\bepsilon^{(j)})} + \nabla_\phi\log \left|J_{T_\phi}(\bepsilon^{(j)})\right|_+ \right],
    \end{split}
\end{equation*}
where $\bepsilon^{(j)} \mathop{\sim}^{\text{i.i.d.}} p_\text{base}(\bepsilon)$ and $\bw_\cS^{(j)} = T_\phi(\bepsilon^{(j)})$. 
In the above expression, by the chain rule, the gradient of the log Jacobian determinant decomposes additively across the $K$ layers: $\nabla_\phi\log|J_{T_\phi}(\bepsilon)|_{+} = \sum_{k=1}^K \nabla_\phi\log|J_{T_k}(\bz_{k-1})|_+$, where 
$\bz_k = T_k(\bz_{k-1})$, $k=1,\ldots,K$, with $\bz_0 = \bepsilon$ and $\bz_K = \bw_\cS$.
In practice, these gradients are computed using automatic differentiation.

In our implementation, we parameterize the transformations $T_1,\ldots,T_K$ using the Masked Autoregressive Flow \parencite[MAF,][]{papamakarios2017masked}. The fundamental architecture of a MAF layer consists of an \textit{affine transformer} and a \textit{masked neural network conditioner}. For a given layer mapping an input vector $\bz=(z_1,\ldots,z_k)^\top$ to an output vector $\bz'=(z_1',\ldots,z_k')^\top$, the affine transformer applies an element-wise affine transformation:
\begin{equation*}
    z_i' = z_i \cdot \exp(a_i) + b_i, \quad, i=1,\ldots,k,
\end{equation*}
where $a_i$ and $b_i$ are the log-scale and translation parameters, respectively. 
The conditioner is a neural network that maps the input $\bz$ to the parameters $(a_1,b_1,\ldots,a_k,b_k)^\top$ in a single forward pass. 
Importantly, the neural network is a masked autoencoder (MADE) with binary masks on the weights, such that $a_i$ and $b_i$ depend only on the preceding input dimensions $\bz_{1:i-1}$.
This autoregressive structure implies that the Jacobian matrix of the transformation is lower-triangular, with diagonal entries exactly equal to $\exp(a_i)$. 
Consequently, the log Jacobian determinant $\log |J_{T_\phi}(\bz)|_+$ simplifies to $\sum_i a_i$.

\section{Simulation of the Hierarchical Model}\label{sec:simulation-model}
The responses $\by_\cT$ in the hierarchical model (\ref{eq:hierarchical-model-subset}) can be efficiently simulated in $O((n\vee k)(\tilde m^2 + m^3))$ via the procedure below. 

\begin{minipage}{0.9\linewidth}
\begin{algorithm}[H]
\caption{Simulating $\by_\cT$}
%\footnotesize
\begin{algorithmic}
\STATE {\textbf{Input}} Inducing points $\bZ=\{\bz_1,\ldots,\bz_{\tilde m}\}$; model parameters $(\bbeta,\sigma^2_\epsilon)$;
\STATE Simulate $\bg_{\bZ} \sim N(\bzero, \bR_\bZ)$ on the inducing points;
\STATE Simulate $\bw_{1:m}\sim N(\bw_{1:m}|\bzero, \bC_{mm})$;
\FOR{$i = m+1$ \TO $k$} 
\STATE Recursively simulate 
    \begin{equation*}
            g_i \sim N(\bR_{\tilde\bw_i,\bZ}\bR_\bZ^{-1}\bg_{\bZ}, 1 - Q(\tilde\bw_i,\tilde\bw_i)), \quad w_i \sim N(h_i + \tau_i g_i, F_i);
    \end{equation*}
    where $\tilde\bw_i=\Lambda^{1/2}_{\bs_i,N(\bs_i)}\bw_{N(\bs_i)}\in\bbR^m$, $h_i = \bB_{\bs_i, N(\bs_i)}\bw_{N(\bs_i)}$, $\tau_i = \tau(\bs_i,N(\bs_i))$, $F_i = C_{\bs_i|N(\bs_i)}$,
    and $Q(\bx,\bx') = \bR_{\bx,\bZ}\bR^{-1}_{\bZ}\bR_{\bZ,\bx'}$ is the Nystr\"om approximation of $R$;
\ENDFOR
\FOR{each $\bu\in\cT\backslash\cS$}
\STATE Independently simulate 
    \begin{equation*}
            g_{\bu} \sim N(\bR_{\tilde\bw_\bu,\bZ}\bR_\bZ^{-1}\bg_{\bZ}, 1 - Q(\tilde\bw_\bu, \tilde\bw_\bu)), \quad w_\bu \sim N(h_\bu + \tau_\bu g_\bu, F_\bu),
    \end{equation*}
    where $\tilde\bw_\bu=\Lambda^{1/2}_{\bu,N(\bu)}\bw_{N(\bu)}\in\bbR^m$, $h_\bu = \bB_{\bu, N(\bu)}\bw_{N(\bu)}$, $\tau_\bu = \tau(\bu,N(\bu))$, and $F_\bu = C_{\bu|N(\bu)}$;
\ENDFOR
\STATE Simulate $\by_\cT \sim N(\bX_\cT\bbeta + \bw_\cT, \sigma^2_\epsilon\bI)$.
\STATE {\textbf{Output}} $\by_\cT$.
\end{algorithmic}
\end{algorithm}
\end{minipage}
\vspace{.2in}

When sampling $g$ in the above procedure, we have used the fact that under the FIC approximation (\ref{eq:FIC}), the conditioning process of $g(\bz)$ given $\bg_\bZ$ is location-wise independent with 
mean $\bR_{\bz,\bZ}\bR_\bZ^{-1}\bg_{\bZ}$ and variance $R(\bz,\bz) - Q(\bz,\bz)$ \parencite{quinonero2005unifying}. 
As a result, given $\bg_\bZ$, $g_i$ is independent of $g_{<i}$.

\section{Evaluation Metrics for Extreme Values}\label{sec:metrics-extreme}
In this section, we explain the metrics used in the main paper to assess the predictive performance on extreme events. We focus on the left tail of the distribution. 
The right tail case can be handled similarly.
Let $\mathcal{U} = \{\bu_1, \dots, \bu_N\}$ be the set of spatial locations in the test dataset, and let $y_{\bu}$ denote the standardized true observation at location $\bu \in \mathcal{U}$. 
Let $F_{\bu}(\cdot)$ be the predictive cumulative distribution function (CDF) generated by the model at location $\bu$, which can be estimated by the empirical CDF of an ensemble of $L$ posterior predictive samples $\{\hat y_{\bu,1}, \hat y_{\bu,2}, \dots, \hat y_{\bu,L}\}$.
Let $c$ be a pre-specified scalar threshold, which is typically set to the left quantile of the standard normal distribution (e.g., $c = -1.64$ for the $5\%$ quantile).

\paragraph*{BSS}

The Brier Score \parencite[BS,][]{glenn1950verification} measures the mean squared difference between the predicted probability of an event and its actual occurrence. For the left-tail extreme event $\{y_{\bu} < c\}$, the predicted probability is $F_{\bu}(c)$, and the Brier Score at threshold $c$ is defined as:
\begin{equation*}
    \text{BS}(c) = \frac{1}{N} \sum_{\bu \in \mathcal{U}} \left( F_{\bu}(c) - I(y_{\bu} < c) \right)^2,
\end{equation*}
where $I(\cdot)$ is the indicator function. 
To quantify the predictive skill relative to a baseline, the Brier Skill Score \parencite[BSS,][]{wilks2011statistical} is computed using the sample climatology as the reference model. 
Let $p = \frac{1}{N} \sum_{\bu \in \mathcal{U}} I(y_{\bu} < c)$ be the climatological base rate of the non-exceedance in the test set. 
The corresponding Brier Score is 
\begin{equation*}
    \text{BS}_{\text{climatology}}(c) = \frac{1}{N} \sum_{\bu \in \mathcal{U}} \left( p - I(y_{\bu} < c) \right)^2 = p(1-p).
\end{equation*}
The BSS is then defined as:
\begin{equation*}
    \text{BSS}(c) = 1 - \frac{\text{BS}(c)}{\text{BS}_{\text{climatology}}(c)}.
\end{equation*}
A positive $\text{BSS}(c)$ indicates that the model demonstrates better predictive skill than the empirical climatology.

\paragraph*{twCRPS}

The Continuous Ranked Probability Score \parencite[CRPS,][]{gneiting2007probabilistic} measures the distance between the predictive CDF and the empirical CDF of the true observation. 
To specifically penalize prediction errors in extreme regions, we employ the threshold-weighted CRPS \parencite[twCRPS,][]{gneiting2011comparing}, which is defined as:
\begin{equation*}
    \text{twCRPS}_{\bu}(c) = \int_{-\infty}^{c} \left( F_{\bu}(z) - I(z \ge y_{\bu}) \right)^2 \,\dif z.
\end{equation*}
Calculating the twCRPS is equivalent to computing the standard CRPS on the threshold-transformed variables.
To be specific, we define the transformed predictive samples as $\hat{y}^\dagger_{\bu,l} = \min(\hat y_{\bu,l}, c)$ for $l = 1, \dots, L$, and the transformed true observation as $y^\dagger_{\bu} = \min(y_{\bu}, c)$. 
The sample-based twCRPS at location $\bu$ can then be calculated as
\begin{equation*}
    \text{twCRPS}_{\bu}(c) = \frac{1}{L} \sum_{l=1}^L |\hat{y}^\dagger_{\bu,l} - y^\dagger_{\bu}| - \frac{1}{2L^2} \sum_{l=1}^L \sum_{j=1}^L |\hat{y}^\dagger_{\bu,l} - \hat{y}^\dagger_{\bu,j}|.
\end{equation*}
The overall score is given by the spatial average: 
$$
\text{twCRPS}(c) = \frac{1}{N} \sum_{\bu \in \mathcal{U}} \text{twCRPS}_{\bu}(c).
$$ 
A lower twCRPS value indicates better predictive performance in the left tail. 

\paragraph*{Extreme CI Coverage and Width}

To assess the reliability and sharpness of the predictive intervals for extreme observations, we calculate the empirical coverage rate and the average width of the $(1-\alpha)$ credible intervals (CIs) conditional on the true observation falling below the threshold $c$.

Let $[L_{\bu}, U_{\bu}]$ denote the $(1-\alpha)$ predictive interval at location $\bu$, constructed using the empirical $\alpha/2$ and $1-\alpha/2$ quantiles of the predictive ensemble $\{\hat y_{\bu,l}\}_{l=1}^L$. 
Let $\mathcal{U}_c = \{\bu \in \mathcal{U}: y_{\bu} < c\}$ be the subset of test locations where the true observations are extreme left-tail values, and let $N_c = |\mathcal{U}_c|$ be the number of such locations. 
The extreme CI coverage and extreme CI width are defined as:
\begin{align*}
    \text{Extreme CI coverage}(c) &= \frac{1}{N_c} \sum_{\bu \in \mathcal{U}_c} I(L_{\bu} \le y_{\bu} \le U_{\bu}), \\
    \text{Extreme CI Width}(c) &= \frac{1}{N_c} \sum_{\bu \in \mathcal{U}_c} (U_{\bu} - L_{\bu}).
\end{align*}
An ideal model should yield an Extreme CI coverage close to the nominal level $(1-\alpha)$ while maintaining a relatively narrow CI width.

\section{Additional Simulation Results}\label{sec:additional-simu}
\begin{figure}[!htb]
    \centering
    \begin{subfigure}{0.35\textwidth}
        \centering
        \includegraphics[width=\linewidth]{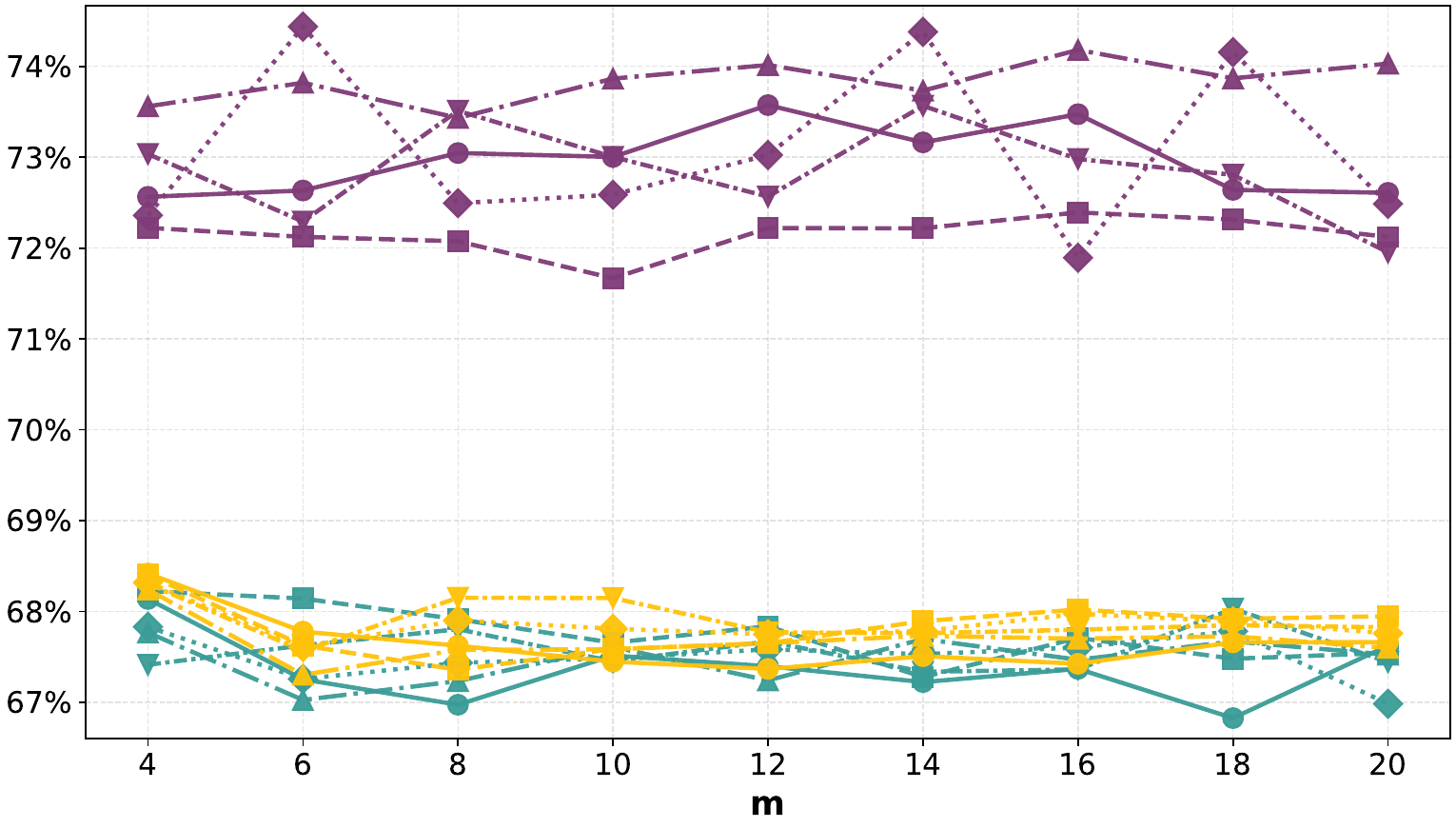}
        \caption{RSR(\%).}
    \end{subfigure} 
    \hspace{.25in}
    \begin{subfigure}{0.35\textwidth}
        \centering
        \includegraphics[width=\linewidth]{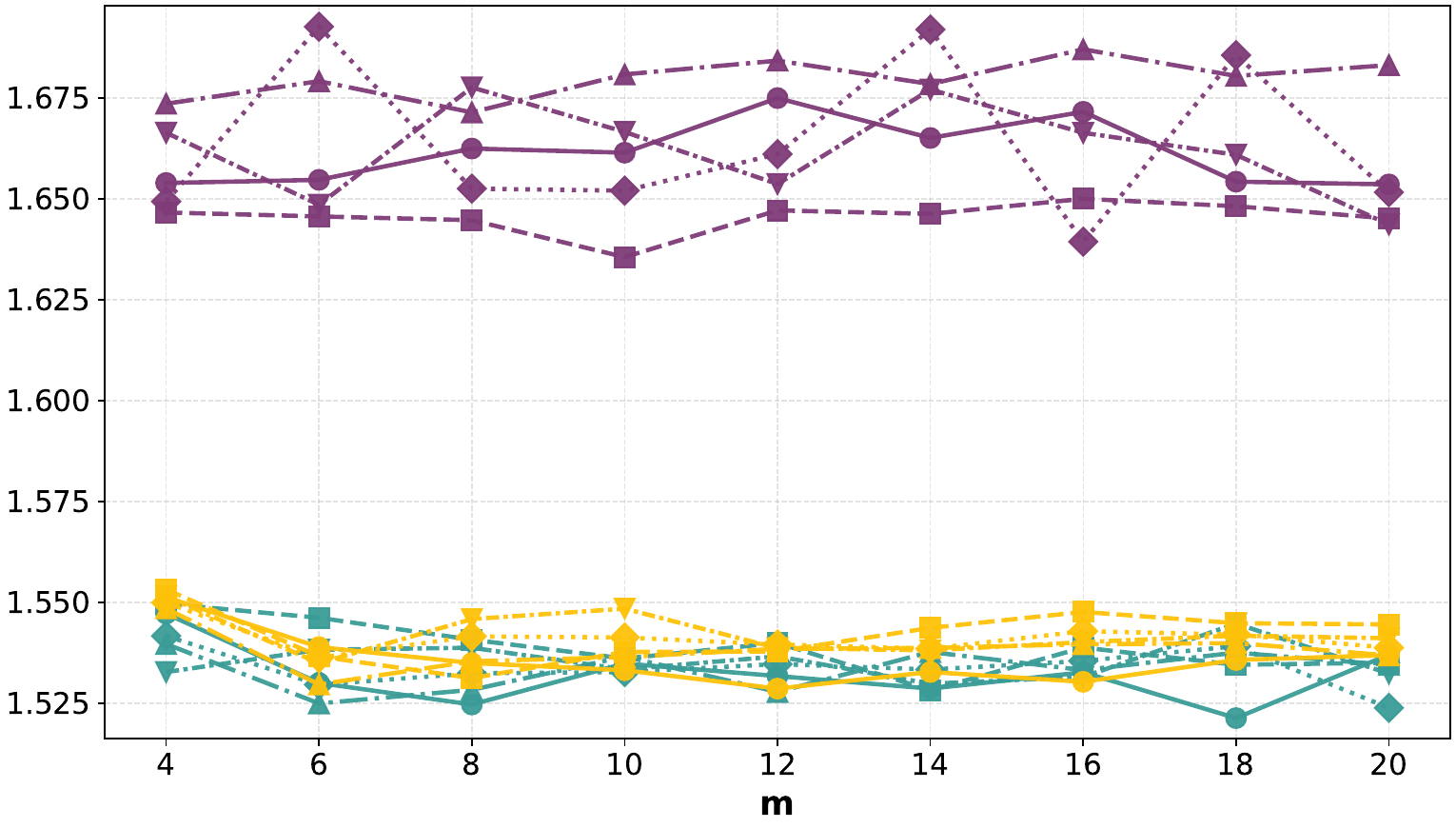}
        \caption{CRPS.}
    \end{subfigure} 

    \vspace{.1in}
    \begin{subfigure}{0.35\textwidth}
        \centering
        \includegraphics[width=\linewidth]{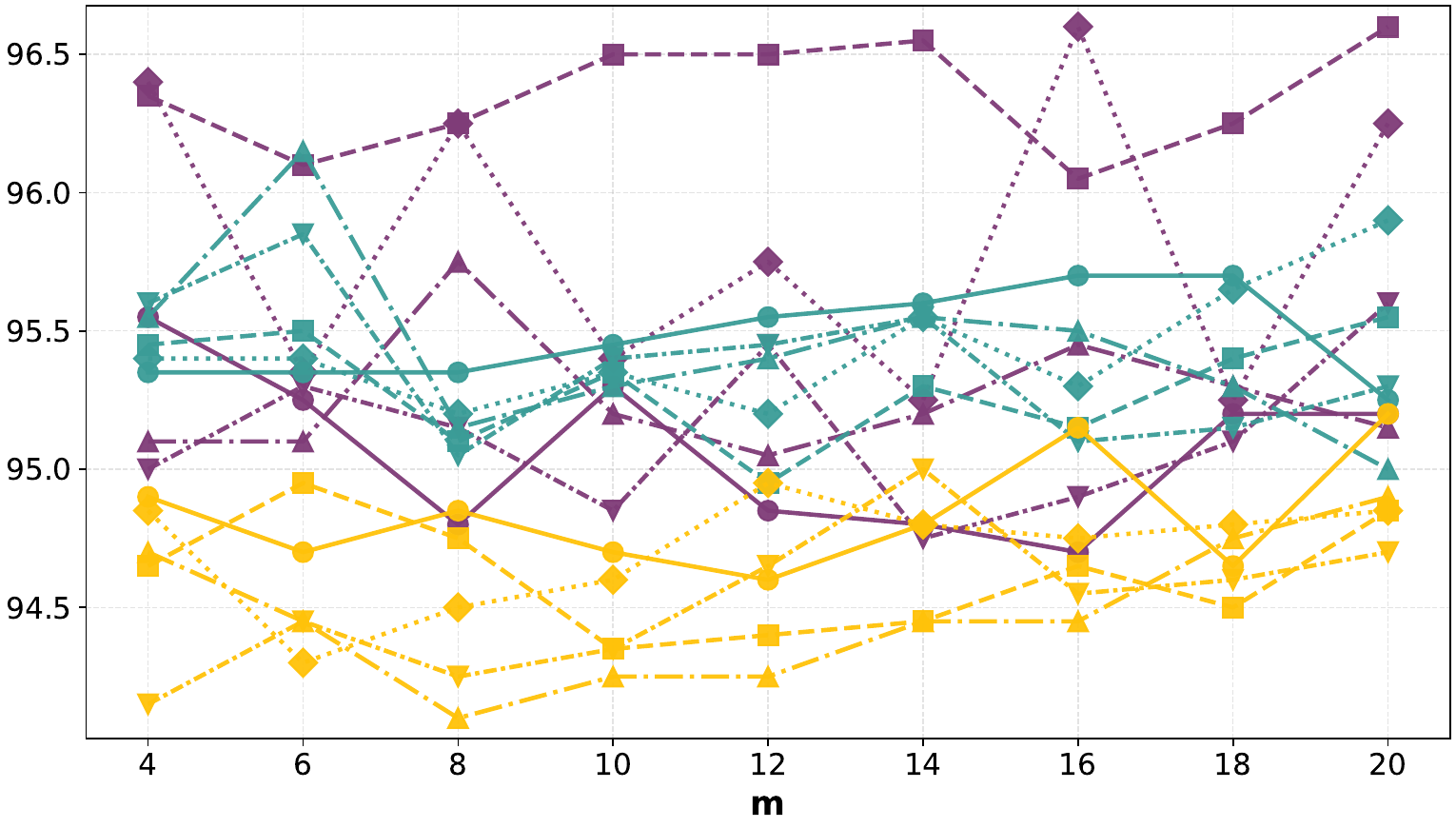}
        \caption{$95\%$ CI coverage rate.}
    \end{subfigure}
    \hspace{.25in}
    \begin{subfigure}{0.35\textwidth}
        \centering
        \includegraphics[width=\linewidth]{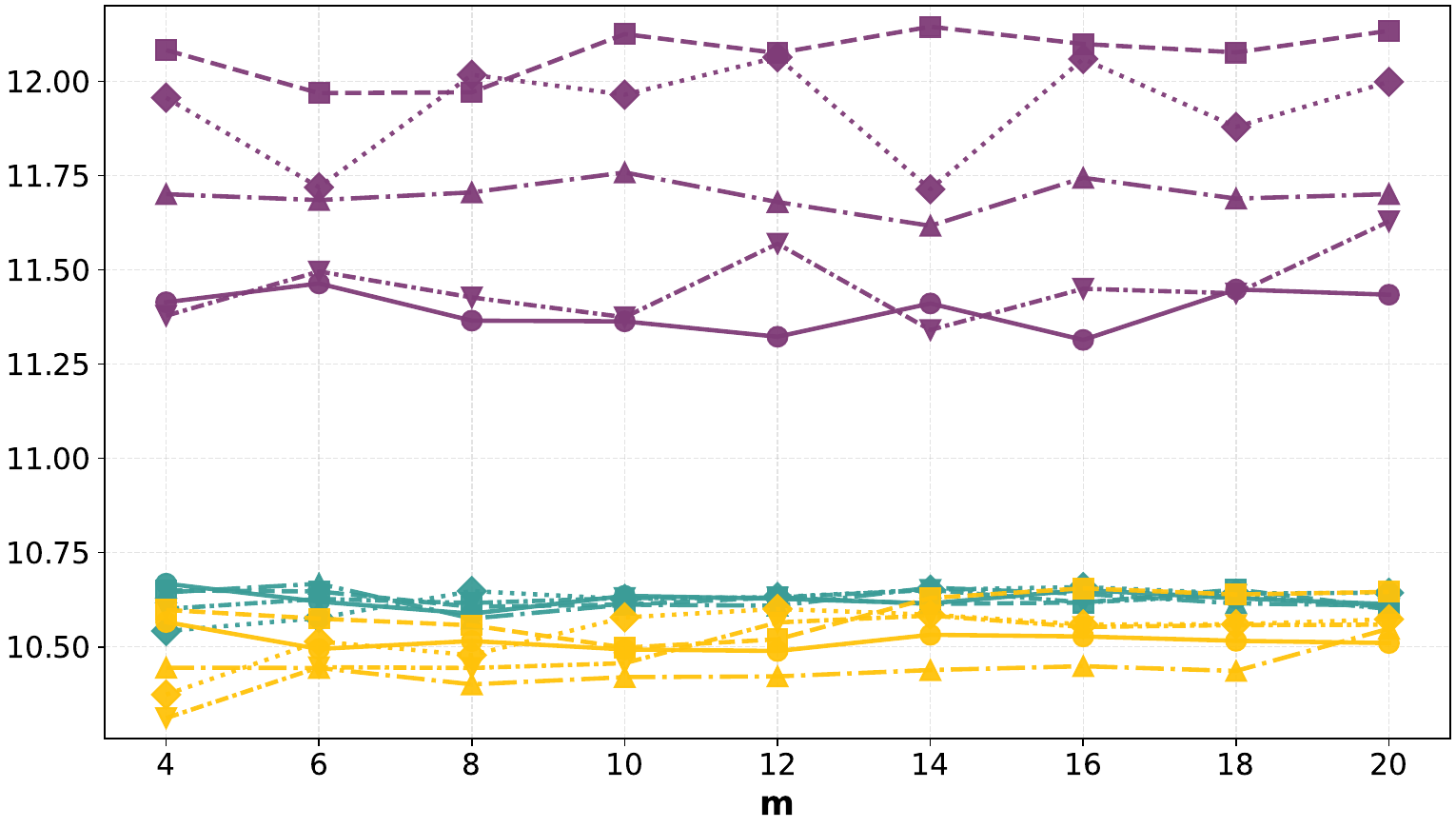}
        \caption{$95\%$ CI width.}
    \end{subfigure}

    \vspace{.1in}
    \includegraphics[height=0.07\textwidth]{figure/methods_orders_legend.pdf}
    \caption{Predictive performance with different orderings and neighbor sizes.}
    \label{fig:result-simu-no-split}
\end{figure}

The following is an extension of the second simulation experiment in Section~\ref{sec:illustrations}.
We create a regular grid of size $50\times 50=2500$ on a square domain $[0,5]^2$. 
The grid points $\{\bs_1,\ldots,\bs_{2500}\}$ are ordered by maximin.
We generate the responses sequentially by
\begin{equation*}
    y(\bs_i) \sim \text{N}(\bB_{\bs_i, N_{10}(\bs)}\by_{N_{10}(\bs_i)} + 3\sin(4\bB_{\bs_i, N_{2}(\bs_i)}\by_{N_{2}(\bs_i)}), \bC_{\bs_i|N_{10}(\bs_i)}),
\end{equation*}
where $N_j(\bs_i)$ are the $j$-nearest neighbors in $\{\bs_1,\ldots,\bs_{i-1}\}$.
A major difference from the second simulation in Section~\ref{sec:illustrations} is that the nearest neighbors are no longer restricted to a reference set.
As a result, all three models: NNGP, NNMP, and NNnGP, are misspecified because they handle the reference points and non-reference points separately.
Figure~\ref{fig:result-simu-no-split} shows the prediction performance of three models with different ordering methods and neighbor sizes.
The NNnGP model achieves comparable performance with NNGP.

\newpage

\printbibliography[title={Supplementary References}]
\end{refsection}

\end{document}